\documentclass{aastex63}

\usepackage{xcolor}
\usepackage{enumerate}
\let\a=\alpha \let\b=\beta \let\g=\gamma \let\d=\delta \let\e=\epsilon
\let\h=\eta  \let\k=\kappa \let\l=\lambda 
\let\s=\sigma \let\t=\tau \let\u=\upsilon
\let\r=\rho
 
\let\w=\omega	   \let\G=\Gamma \let\D=\Delta  \let\L=\Lambda
\let\S=\Sigma \let\W=\Omega
\let \x = \chi \let \z = \xi

\def\be{\begin{eqnarray}}
\def\ee{\end{eqnarray}}
\def\ba{\begin{array}}
\def\ea{\end{array}}

\def \yr {\mathrm{yr}}

\def \M  {\mathscr{M}}
\def \K {\mathrm{K}}

\def \earth {\oplus}
\def \S {Section }
\def \Per {\mathrm{Per}}

\def \Prob {\mathrm{P}}

\shorttitle{Peas in a Pod}
\graphicspath{{./}{figures/}}

\begin{document}

\title{Peas in a Pod? Radius correlations in Kepler multi-planet systems}

%\correspondingauthor{...}
\email{lena@ias.edu, tremaine@ias.edu}

\author[0000-0001-8986-5403]{Lena Murchikova}, \author[0000-0002-0278-7180]{Scott Tremaine}
\affiliation{Institute for Advanced Study \\
1 Einstein Drive \\
Princeton, NJ 08540, USA}

%\nocollaboration{2}

\begin{abstract}

%We address the claim of \cite{2018AJ....155...48W} that the radii of adjacent planets in {\it Kepler} multi-planet systems are correlated. We use a simple toy model to show that this kind of apparent correlation can arise from observational selection effects even if the radii are chosen at random from a universal distribution. We also comment on the validity of a commonly used correction that is used to estimate intrinsic planet occurrence rates, based on weighting planets by the inverse of their detectability.

We address the claim of \cite{2018AJ....155...48W} that the radii of adjacent planets in {\it Kepler} multi-planet systems are correlated. We explore two simple toy models---in the first the radii of the planets are chosen at random from a single universal distribution, and in the second we postulate several types of system with distinct radius distributions. We show that an apparent correlation between the radii of adjacent planets similar to the one reported by \cite{2018AJ....155...48W} can arise in both models. In addition the second model fits the radius and  signal-to-noise distribution of the observed planets. We also comment on the validity of a commonly used correction that is used to estimate intrinsic planet occurrence rates, based on weighting planets by the inverse of their detectability.

\end{abstract}


\section{Introduction} \label{sec:intro}

\citet[][hereafter W18]{2018AJ....155...48W} reported correlations in the characteristics of neighbouring planets in multi-planet systems -- a phenomenon they called ``peas in a pod". One of these correlations is in the planetary radii: as stated by W18, ``each planet is more likely to be the size of its neighbor than a size drawn at random from the distribution of observed planet sizes". Here we show that the correlation found by W18 can be largely produced by observational selection effects and/or system-to-system variations in the distribution of planetary radii.

Before beginning it is worthwhile to describe the relevant hypotheses more carefully. There are (at least) three logical possibilities:

\begin{enumerate}[(i)]
    
    \item ``Planets don't know anything". More formally, this hypothesis states that the planet-formation process is universal, in the sense that the probability that a planet has radius in a small range $R_p\to R_p+dR_p$ is a universal function, independent of the properties of the host system or the presence or properties of other planets. 
    
    \item ``Planets know about the system they formed in". This is the hypothesis that the probability distribution of radii in a host system labeled by $\alpha$ is some function $p(R_p|\alpha,\mathrm{Per})$ where Per is the orbital period. Thus the radius distribution depends on the properties of the system and the planet's location in it, but not on the presence of other planets.
    
    \item ``Planets know about other planets". This is the hypothesis that the probability distribution of radii in an $N$-planet system is a function $p(R_{p1},\ldots,R_{pN})$ that cannot be separated into a product $\prod_{j=1}^N p(R_{pj})$; in other words the correlation coefficient between the distribution of radii of different planets in the same system is non-zero \citep[e.g.,][]{2018MNRAS.473..784K,2018AJ....156...24M,2019MNRAS.489.3162S,2019MNRAS.490.4575H,2020arXiv200311098G}.
    
\end{enumerate}

W18 do not state explicitly whether they are arguing for hypothesis (ii) or (iii). Either hypothesis can generate correlations of the kind they observe -- for example, if the radius probability distribution in system $\alpha$ is $p(R|\alpha)=\delta(R-R_\alpha)$ with $R_\alpha$ varying from system to system, then all planets in a given system have the same radius and the correlation coefficient measured by W18 will be unity. The main point of this paper, however, is that the correlations they observe can be produced in some respects by hypothesis (i) and in most respects by a simple version of hypothesis (ii). We stress that we consider only the correlation between the radii of adjacent planets and not the other correlations, some of which are discussed by W18. In particular, the full set of parameters also includes periods. In the simplest cases, e.g., hypotheses (i) and (ii), the period and radius distributions are separable.

The analysis of W18 is based on the ``CKS multis" or ``CKSM" catalog -- a subset of the catalog from the California-Kepler Survey (CKS) containing host stars with more than one detected planet. The CKSM catalog contains 909 planets in 355 multi-planet systems, all with measured signal-to-noise ratio $\mathrm{SNR} \geq 10$ (see eq.\ \ref{eq:snr}). A better procedure would be to work directly with the final (DR25) planet candidate catalog from Kepler and to model the detection efficiency, but since our focus is on the statistical analysis we prefer to use the same sample as W18. An important feature of the CKSM catalog is that it contains only systems with detected planets. Such catalogs are much more difficult to correct for selection effects than catalogs compiled with uniform selection criteria that are independent of whether or not a planet is detected.

Our main arguments concerning hypothesis (i) are contained in Sections \ref{sec:creation} and \ref{sec:corr1}. In \S\ref{sec:boot} we examine the statistical tests used in this context by  \cite{2018AJ....155...48W}, \cite{2017ApJ...849L..33M}, \cite{2019arXiv190805833W} and others to correct for detection biases. We comment on the intrinsic planet radius distribution and the observed SNR distribution and on correlations between the radii of adjacent large planets in Sections \ref{sec:gamma} and \ref{sec:large}. \S \ref{sec.snr.fit} contains our main arguments concerning hypothesis (ii). We summarize in \S\ref{sec:sum}. The goal of this work is not to provide a state-of-the-art model for the observed data, but rather to discuss the methods that are used to analyze such data.

\bigskip

\section{``Planets don't know anything''. Creating mock CKS systems, detectability criteria, and correlation measures}
\label{sec:creation}

We create a mock universe in which planets in a given multi-planet system are not correlated in radius. In particular, we make the strong assumption that the distribution of planets in radius in the mock universe is independent of the properties of the stars, the environment, the orbital period, and the properties of the neighboring planets, that is, the probability that a given planet has radius in the range $R_p\to R_p+dR_p$ is a universal function $p(R_p)dR_p$. 

For simplicity we postulate that in our mock universe the universal distribution of radii is a power law,  
\be{}p(R_p)dR_p \sim R_p^{-\gamma} dR_p
\label{eq:gamma}
\ee
between $R_\mathrm{min}$ and $R_\mathrm{max}$. We set $R_\mathrm{min}$ and $R_\mathrm{max}$ to be the smallest and largest planet radii observed in the CKSM sample, $0.34\,R_\earth$ and $13.11\,R_\earth$. For the power-law exponent we choose $\gamma=4$. The resulting radius distribution is unrealistic in several respects---it does not produce the gap in the radius distribution centered on $2R_\oplus$ that is observed at short orbital periods, and it produces too many small planets compared to the number of large ones---but it provides a good illustration of the effects of observational selection on correlations. We discuss the reasons for the choice of this model in Section \ref{sec:gamma}. This distribution of planets is the ``intrinsic mock" distribution.

The CKSM catalog contains 909 planets. Each planet is characterized by a radius $R_{pj}$, an orbital period $\mathrm{Per}_j$, and a host star $(\star_j)$ with properties such as radius $R_{\star j}$ and mass $M_{\star j}$ that help to determine the detectability of the planet. We call the combination of host star and orbital period the ``position" of planet $j$. In our mock universe we will randomize the radius of the planet according to various prescriptions but will keep the set of positions fixed, i.e., the same as in the CKSM catalog (other choices of how to randomize are possible; see footnote \ref{foot:two}).

In order to populate the CKSM systems we create a uniformly spaced array of 5000 planetary radii $R_{pj}$ between $R_\mathrm{min}$ and $R_\mathrm{max}$.  Each radius in the array is assigned a weight $(w_j)$ such that $w_j = R_{pj}^{-4} /\sum\limits_i R_{pi}^{-4}.$ We then sample from this weighted distribution to populate the CKSM positions one by one with planets. For a given position, we keep drawing planets at random until a planet is detectable based on the signal-to-noise ratio (SNR) criterion 
\citep{2018AJ....155...48W,2017ApJ...849L..33M,2019arXiv190702074Z,2019arXiv190805833W}: 
\be
    \mathrm{SNR}[\star ;\mathrm{Per}, R_p]=\frac{(R_p / R_\star)^2}{{\mathrm{CDPP}_\star}_{6hr}\sqrt{6\,\mathrm{hr} /T(\star ; \mathrm{Per})}} \, \sqrt{\frac{3.5 \, \yr}{\mathrm{Per}}} \geq 10,
    \label{eq:snr}
\ee
with transit duration estimated as
\be
    T(\star ;\mathrm{Per}) = 13 \mathrm{\,hr} \left(\frac{\mathrm{Per}}{1\yr}\right)^{1/3} \left(\frac{\r_{\star}}{\r_{\sun}}\right)^{-1/3}.
\ee
Here $(R_p / R_\star)^2$ is the fraction of the star's area blocked by the planet during transit, $ (3.5 \, \yr)/\mathrm{Per}$ is the number of transits over the lifetime of the {\it Kepler} mission, and $\r_{\star}$ and $\r_{\sun}$ are the mean densities of the star and the Sun respectively. Finally ${\mathrm{CDPP}_{\star}}_{6hr}\sqrt{6\,\mathrm{hr}/T}$ is a measure of the detectability of the planet over the duration of the transit \citep{2012PASP..124.1279C}, in which $\mathrm{CDPP}_{\star 6hr}$ (Combined Differential Photon Precision) is a measure of the photometric noise of the star tabulated in units of parts per million (ppm). For a star with $\mathrm{CDPP}_{\star 6hr}=10$ ppm a 6-hour transit of depth 10 ppm would have $\mathrm{SNR}=1$ (assuming a box-shaped transit, no gaps in the data, etc.; future studies can and should use a more careful detection model). The sample of planets produced in this way is the ``observed mock" sample.

Following W18, we estimated the strength of the apparent correlation between adjacent planetary radii using the Pearson correlation coefficient $r$ in $\log R_p$, as calculated with the package \verb|scipy.stats.pearsonr|.
Pearson's $r$ measures a linear relationship between two datasets. The values of $r$ are between $-1$ and $+1$ with $0$ implying no correlation and $\pm 1$ implying an exact linear relationship. The significance level or $p$-value of Pearson's $r$ depends on the underlying distribution of planetary radii unless the number of points $N$ is sufficiently large (typically $N\gtrsim 500$). Since our data set has an $N$ of almost exactly 500\footnote{W18 had $N=504$ compared to our $N= 508$. We believe that the difference is due to marginal rounding differences.}, the significance levels of Pearson's $r$ may be biased. For this reason, we have also repeated all of our analyses using the Spearman correlation coefficient $r_s$, which is a non-parametric estimator and thus more robust. Spearman's $r_s$ is calculated with the package \verb|scipy.stats.spearmanr|.

\bigskip

\section{Mock universe versus CKS catalog}

\label{sec:corr1}

\begin{figure*}
\vspace{-0.0cm}
\centering
\begin{tabular}{cc}
\includegraphics[width=0.45\columnwidth]{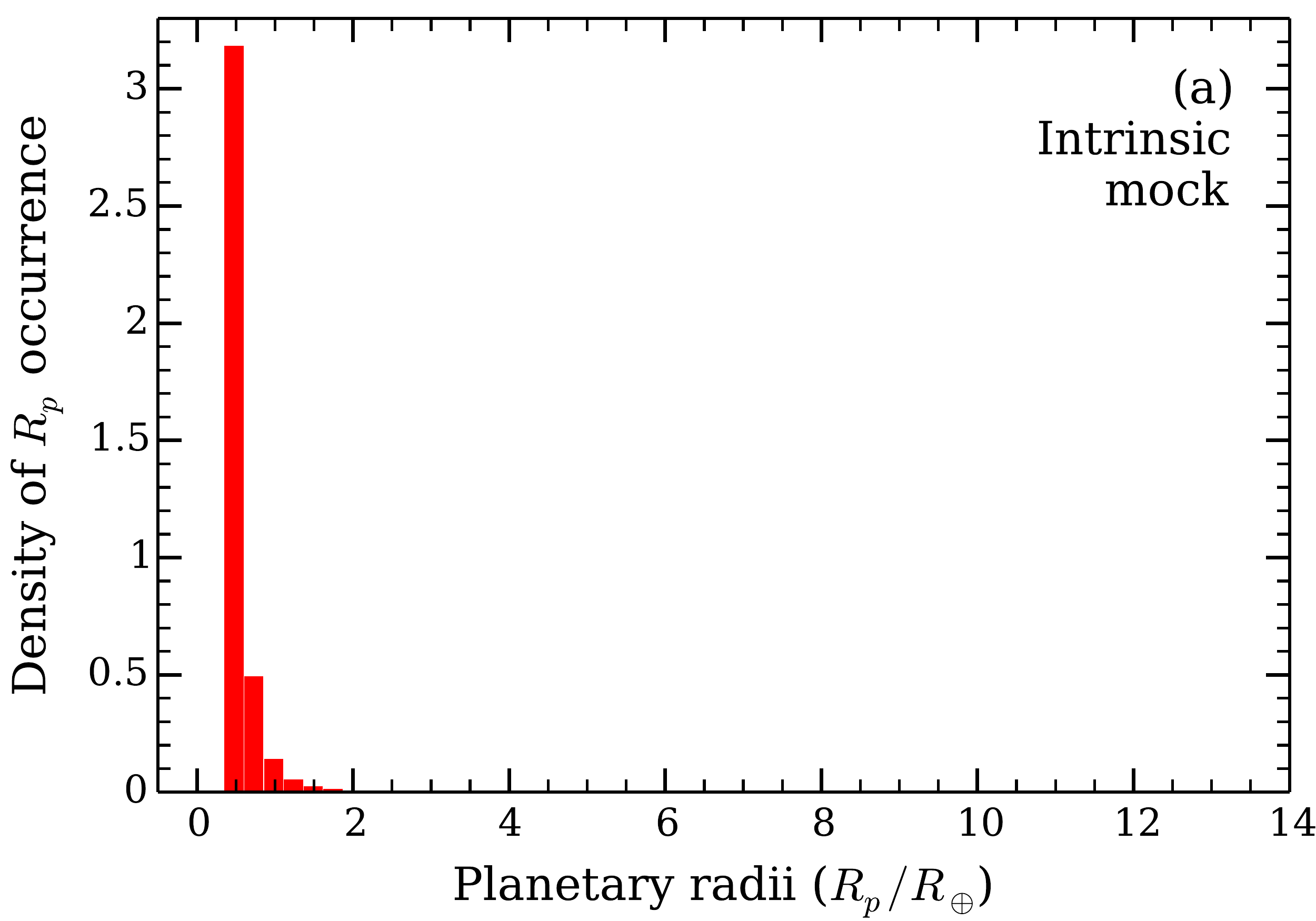} & 
\\
\includegraphics[width=0.45\columnwidth]{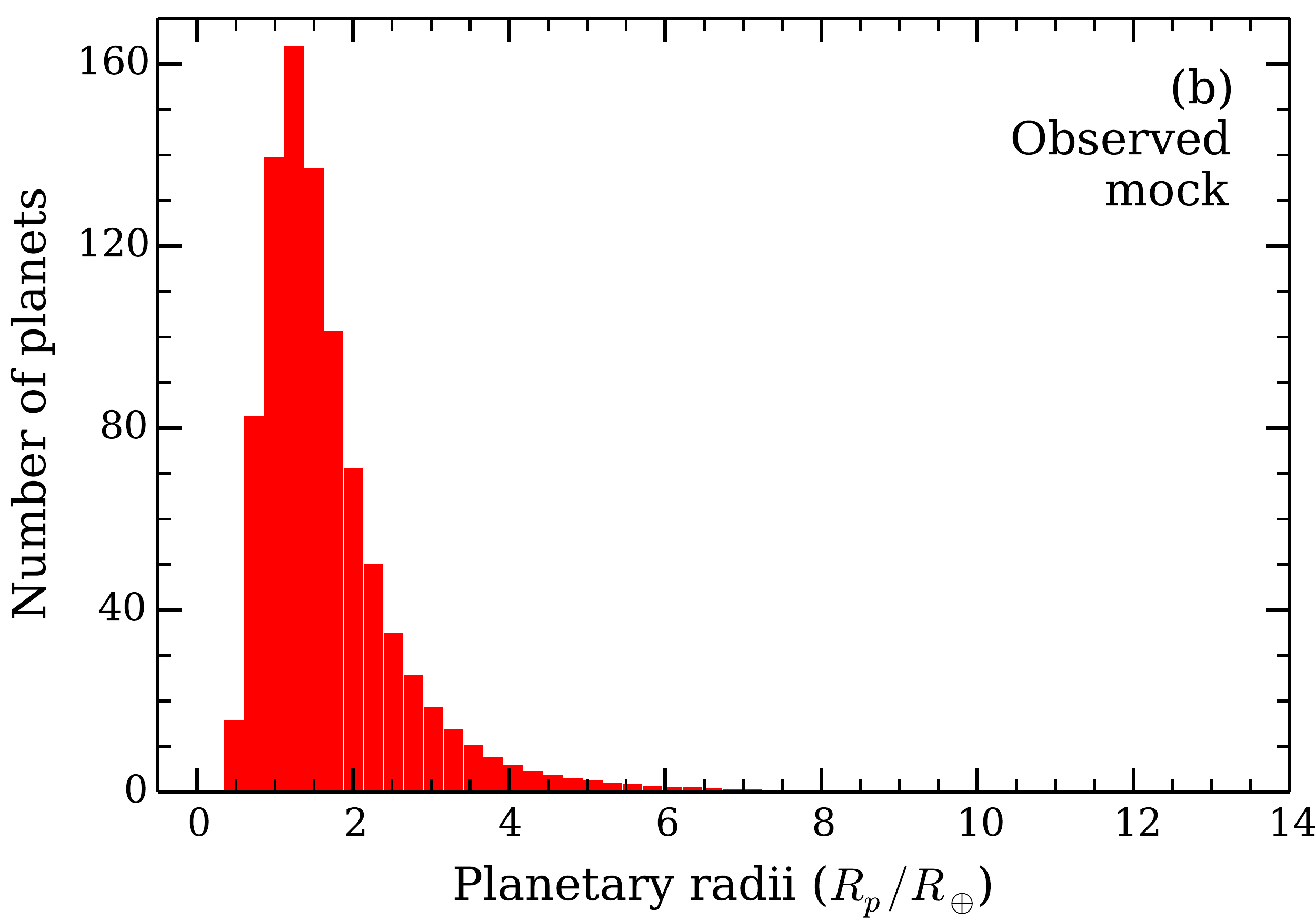} &
\includegraphics[width=0.45\columnwidth]{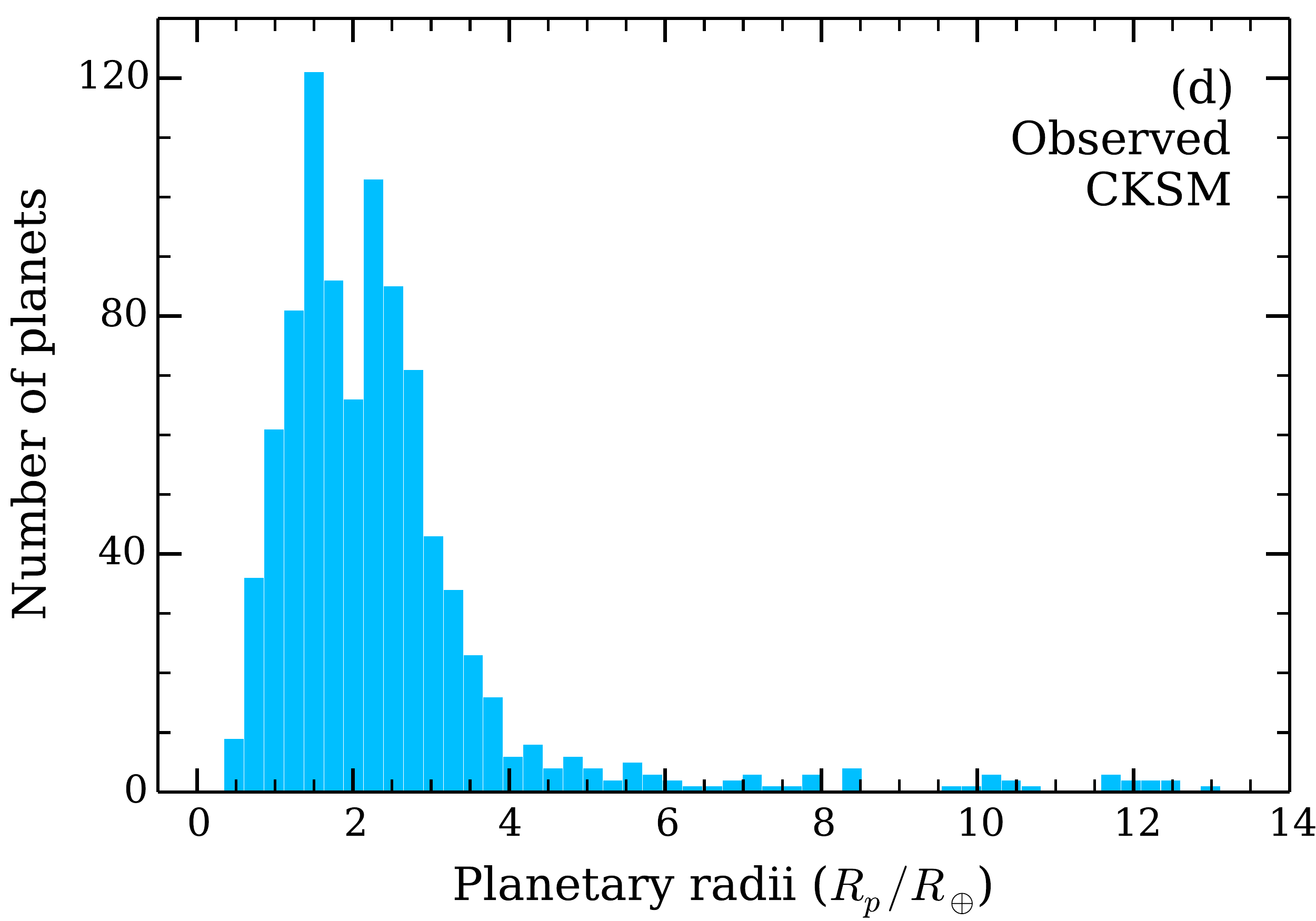}
\\
\includegraphics[width=0.45\columnwidth]{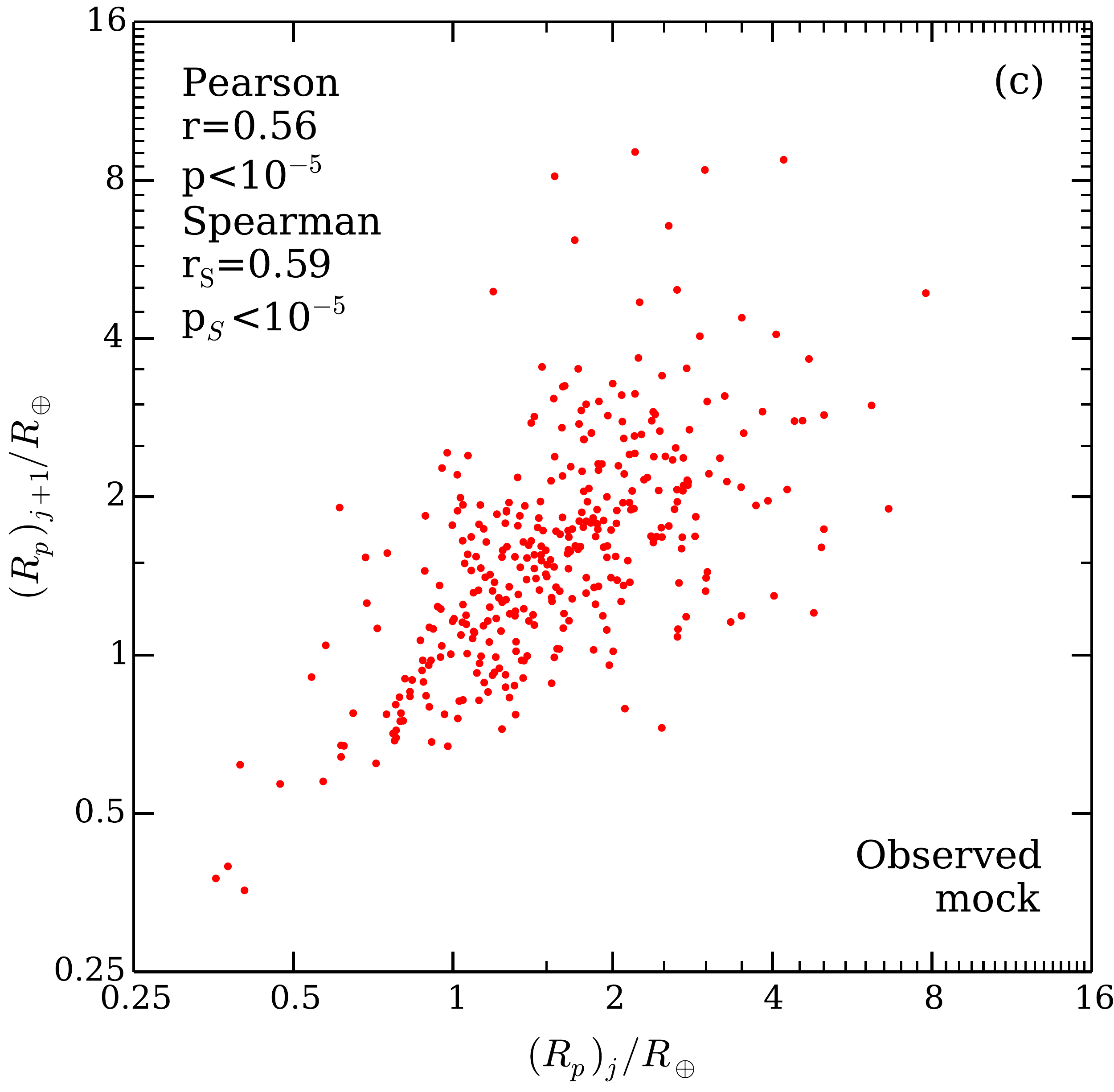}& 
\includegraphics[width=0.45\columnwidth]{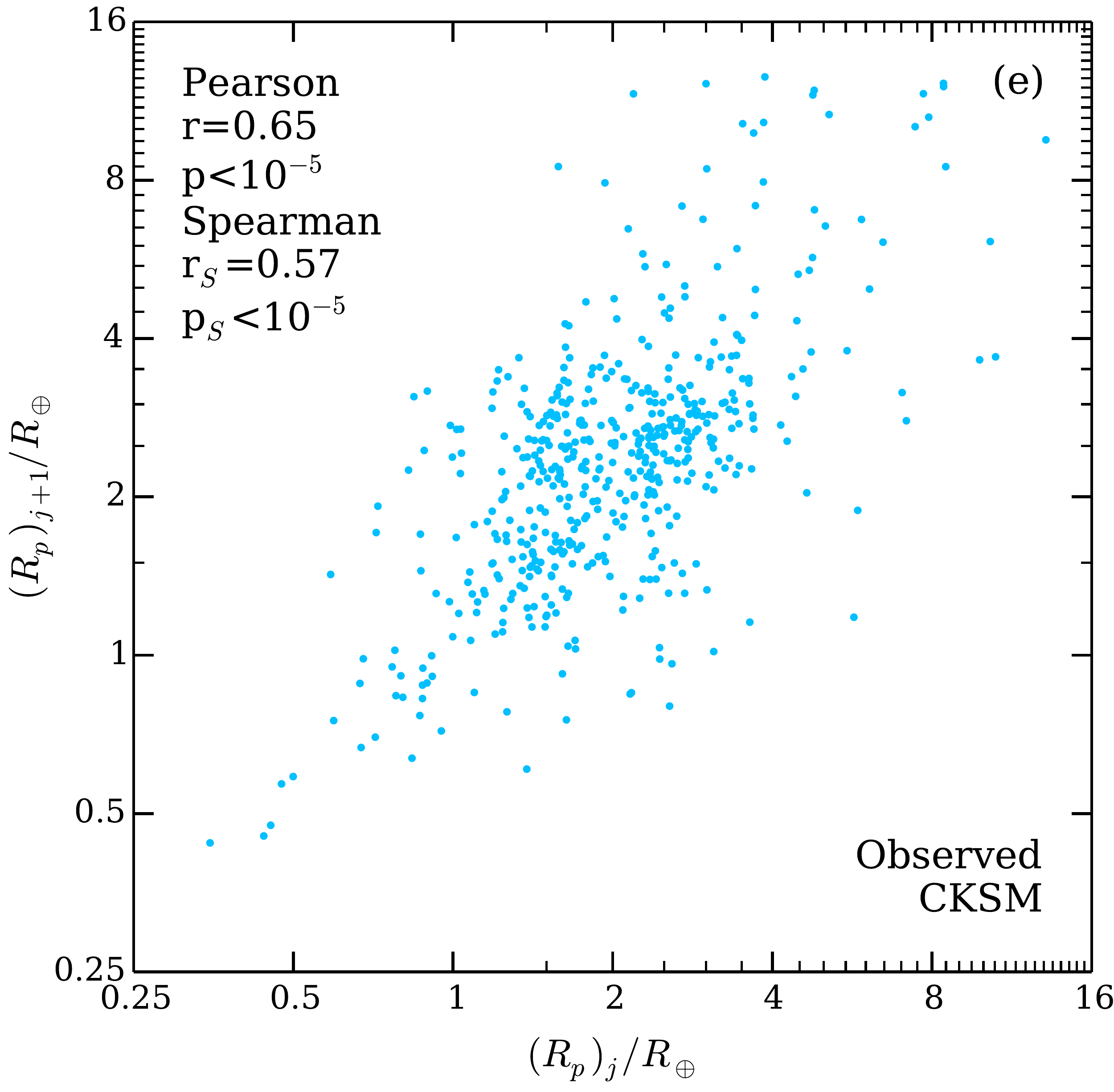} 
\end{tabular}
\caption {Observations in the mock universe compared to the CKSM sample. Red plots belong to the mock universe and blue to the observed planets. (a) The intrinsic mock distribution of planetary radii. The corresponding CKSM panel is empty, because we do not know the underlying distribution of radii in the real universe. (b) The distribution of radii of the observed mock planets, averaged over 5000 Monte-Carlo realizations. For each of the 909 available $(\star_i, \mathrm{Per}_j)$ positions, we draw a radius at random from the intrinsic mock distribution of radii, apply the SNR cut of equation (\ref{eq:snr}), and repeat until the planet survives the cut. (c) The relation between the radii of adjacent planets $R_{pj}$ and $R_{p,j+1}$ in a typical Monte-Carlo realization of mock planets as described for panel (b). The Pearson/Spearman $r$ and $p$ are averages and medians, respectively, of 5000 Monte-Carlo realizations.
(d) and (e) The distribution of radii and the relation between the radii of adjacent planets in the CKSM sample. In panels (c) and (e) we applied the swapping criterion of W18, i.e., a pair of adjacent planets is plotted only if the planets would still be detectable if their positions were swapped.}
\label{distributions}
\end{figure*}

In the figures, we present the plots related to our mock universe adjacent to the ones obtained from the true CKSM planets. The mock data are presented in red and the true data in blue. Figure \ref{distributions}a shows the intrinsic mock distribution of radii, that is, the distribution of the planets in radius as they are created in our mock universe. Figure \ref{distributions}b shows the mean observed mock distribution, i.e., the distribution observed after applying the signal-to-noise cut (\ref{eq:snr}), averaged over 5000 Monte-Carlo realizations. Figure \ref{distributions}d shows the corresponding distribution of radii in the CKSM sample. Apart from the dip in the CKSM distribution at $2R_\oplus$ the two distributions look qualitatively similar. %, which justifies our use for a power law with exponent $\gamma=4$ for the intrinsic radius distribution in the mock universe. 
The dip is real, and is even more prominent in the distribution of planets around single stars in the CKS  \citep{2017AJ....154..109F} and for planets with planets with periods between 10 and 100 days, but we have not attempted to reproduce it in the present mock catalog, since our focus is on simple models and not on careful modeling of the observed data (a mock catalog that does reproduce the dip is described in Section \ref{sec.snr.fit}). We discuss the reasons for choosing a power law with exponent $\gamma=4$ for the intrinsic radius distribution in the mock universe in Section \ref{sec:gamma}.

Figures \ref{distributions}c and \ref{distributions}e contain scatter plots of the radii of adjacent planets in the mock data and real data, analogous to Figure 2 of W18. In preparing these plots, we applied the swapping criterion of W18, i.e., a pair of adjacent planets $(R_{pj},R_{p,j+1})$ is plotted only if the planets would still be detectable if their positions were swapped, $\mathrm{Per}_{j} \leftrightarrow \mathrm{Per}_{j+1}$. The observations, in the panel on the right, show a strong correlation, with Pearson and Spearman correlation coefficients $r=0.65$ and $r_s=0.57$, and $p$-values much less than $10^{-5}$, a conclusion already reached by W18. However, our mock samples on the left \emph{also} show a strong correlation in the adjacent planet radii -- over $5000$ Monte-Carlo realizations the mean Pearson $r=0.56$, the mean Spearman $r_s=0.59$, and the corresponding median $p$-values are much less than $10^{-5}.$

A possible concern with this model is that it implies that the number of undetected planets in some systems is much larger than the number of detected ones, which may lead to gravitational instability. However, systems of planets on nearly circular, nearly coplanar orbits are typically stable if the separation between planets exceeds about 10 Hill radii \citep{2015ApJ...807...44P}. For Earth-mass planets the Hill radius $a(M_\oplus/3M_\odot)^{1/3}$ is 0.01 times the semimajor axis $a$. Thus up to 15 Earth-mass planets could be found between periods of 1 and 10 days without violating this stability constraint. If the radii of these additional planets are smaller than the detection threshold they would be undetected.

We conclude that even if the distribution of planet radii is independent of the properties of the host system and the properties of adjacent planets [hypothesis (i) of the Introduction], observational selection effects can introduce the correlations found in W18. The physical reason for this conclusion was described by \citet{2019arXiv190702074Z}: if the $\mathrm{CDPP}_{\star 6hr}$ or stellar radius $R_\star$ is large, or the transit duration $T(\star;\mathrm{Per})$ is small, then according to equation (\ref{eq:snr}) all planets in a multi-planet system must have large radii to survive the SNR cut in the CKSM catalog. On the other hand if $\mathrm{CDPP}_{\star 6hr}$ or $R_\star$ is small, or $T(\star;\mathrm{Per})$ is large, then both large and small planets can be detected, but the steep slope of the radius distribution means that most planets in a multi-planet system will be small.

% As the philosopher Karl Popper pointed out in ``The Logic of Scientific Discovery'', there is an fundamental asymmetry between verification and falsification. To verify a rule one has to check every instance of evidence in existence, but to falsify a rule it only takes one counter-example. We believe that although our $1/r^4$ distribution is pretty extreme it provides such a counter-example.

\bigskip

\section{Bootstrap and detectability weighting}\label{sec:boot}

\begin{figure*}
\vspace{-0.0cm}
\centering
\begin{tabular}{cc}
\includegraphics[width=0.45\columnwidth]{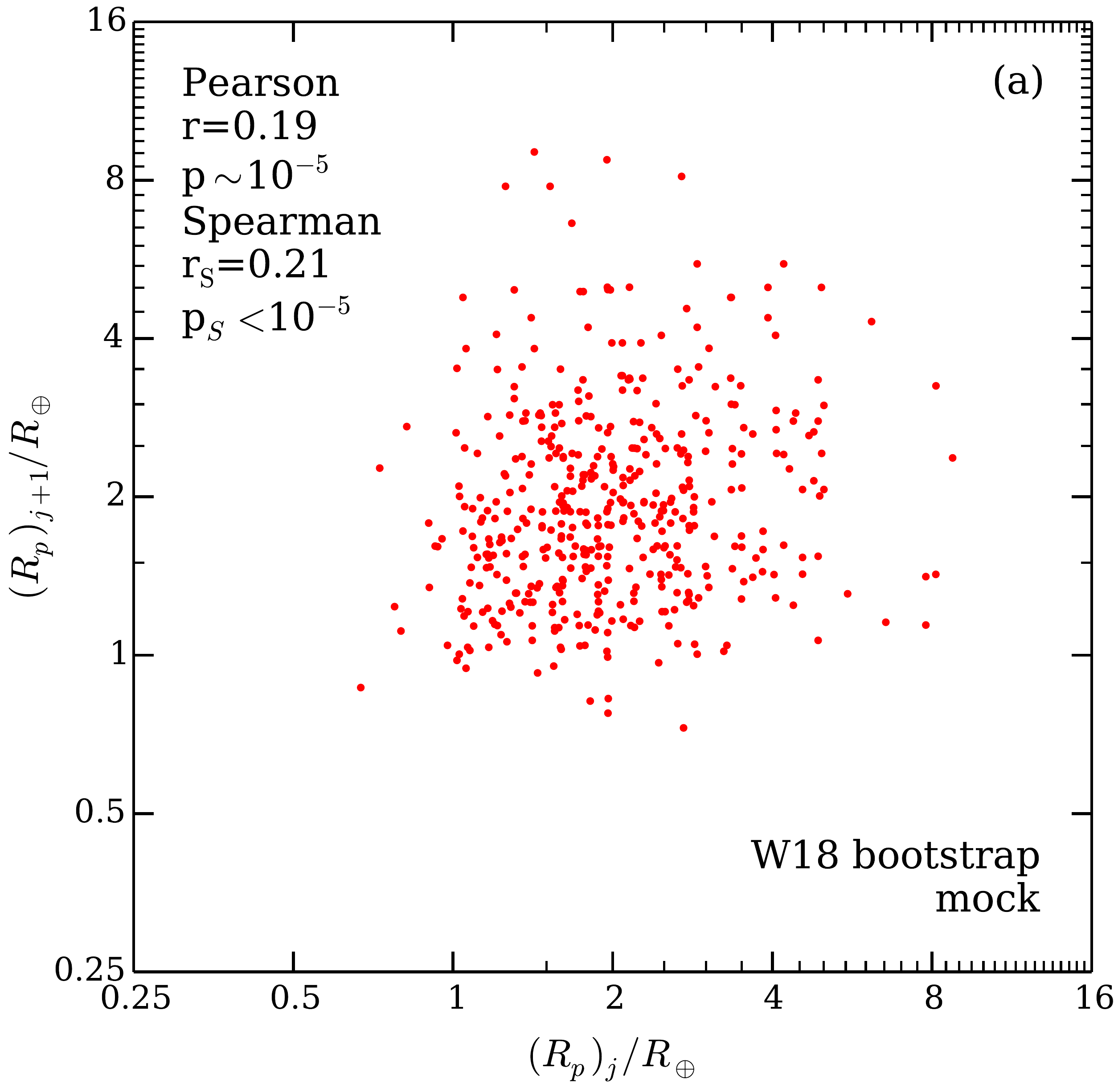} &
\includegraphics[width=0.45\columnwidth]{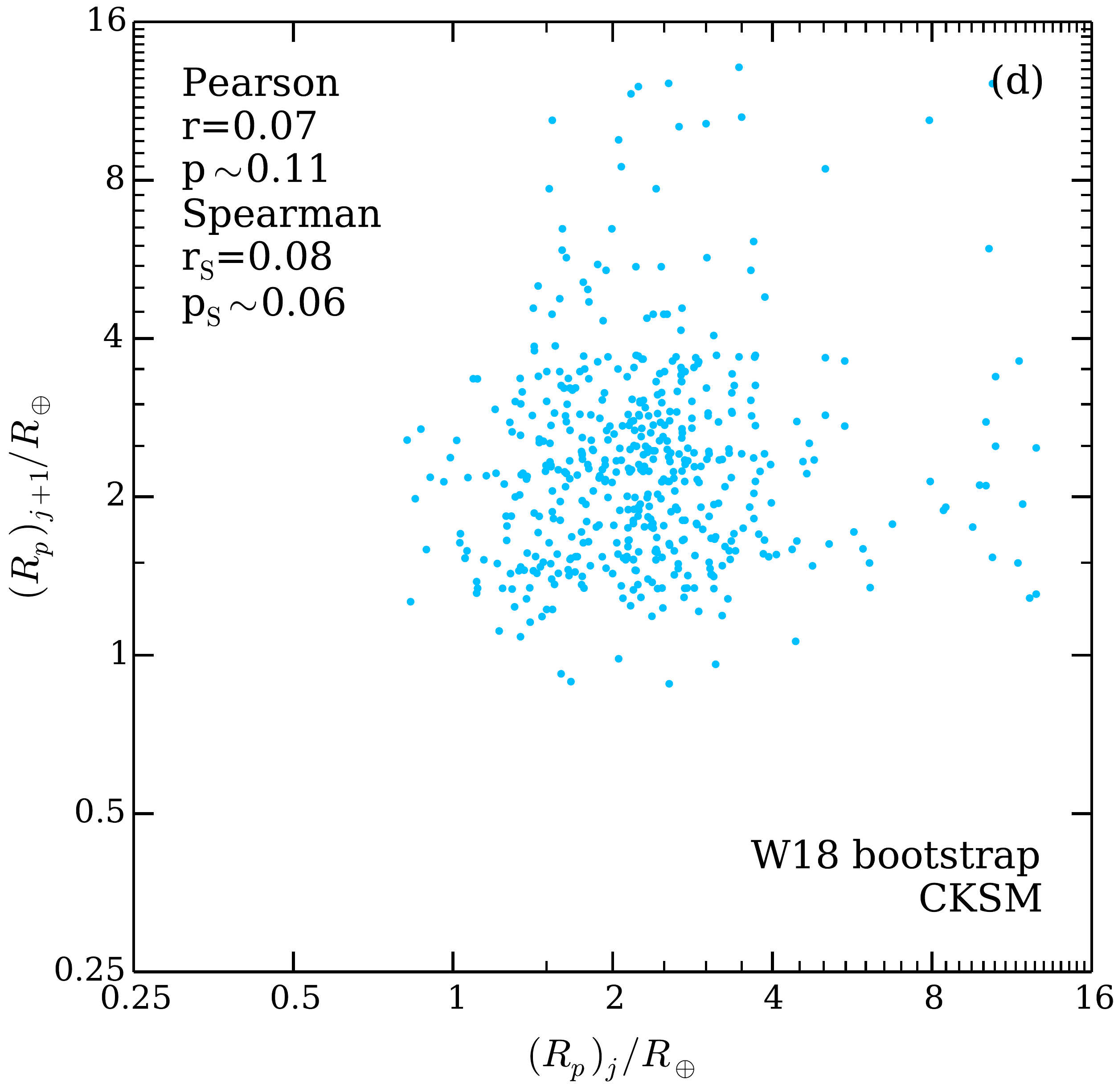}
\\
\includegraphics[width=0.45\columnwidth]{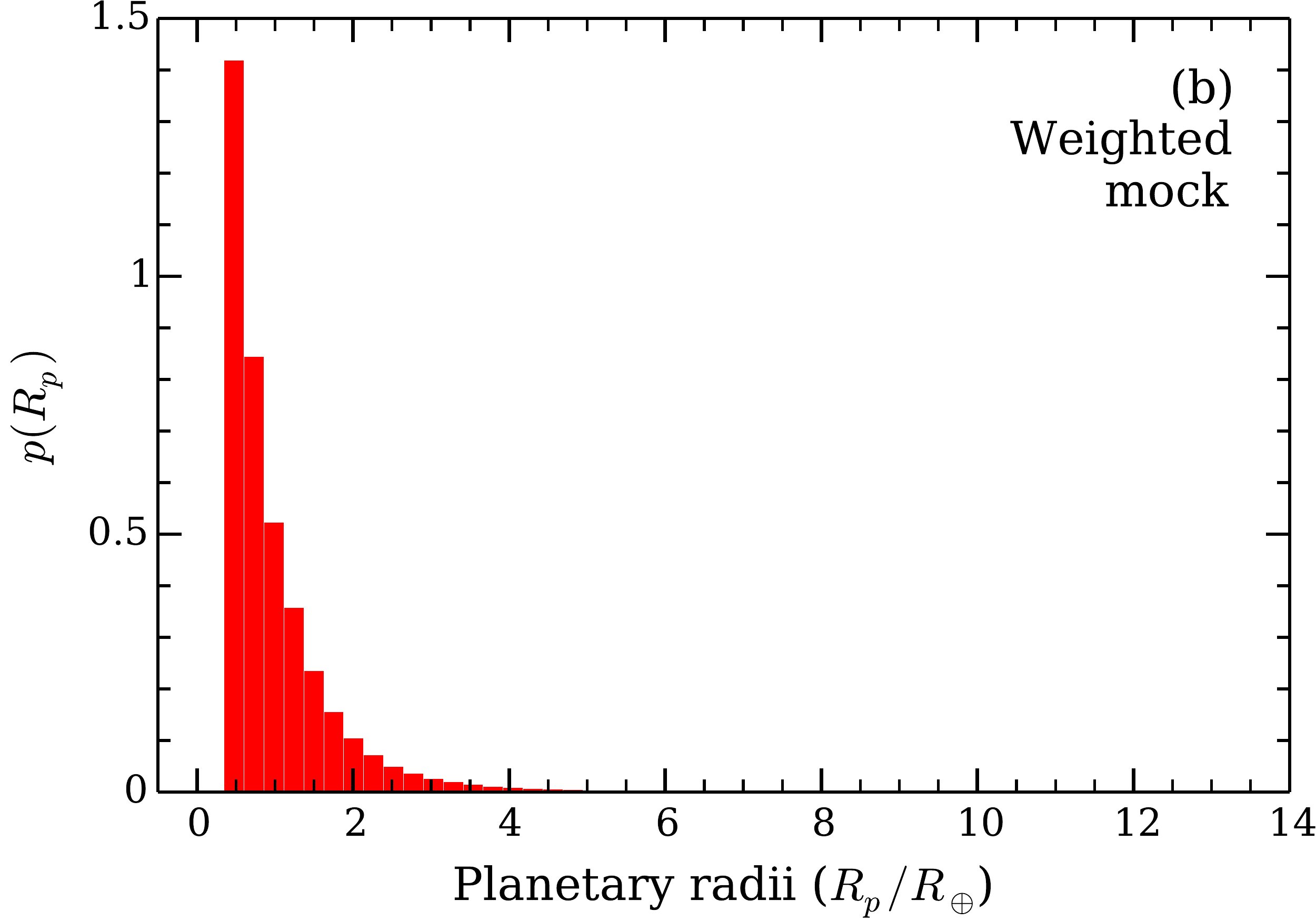} &
\includegraphics[width=0.45\columnwidth]{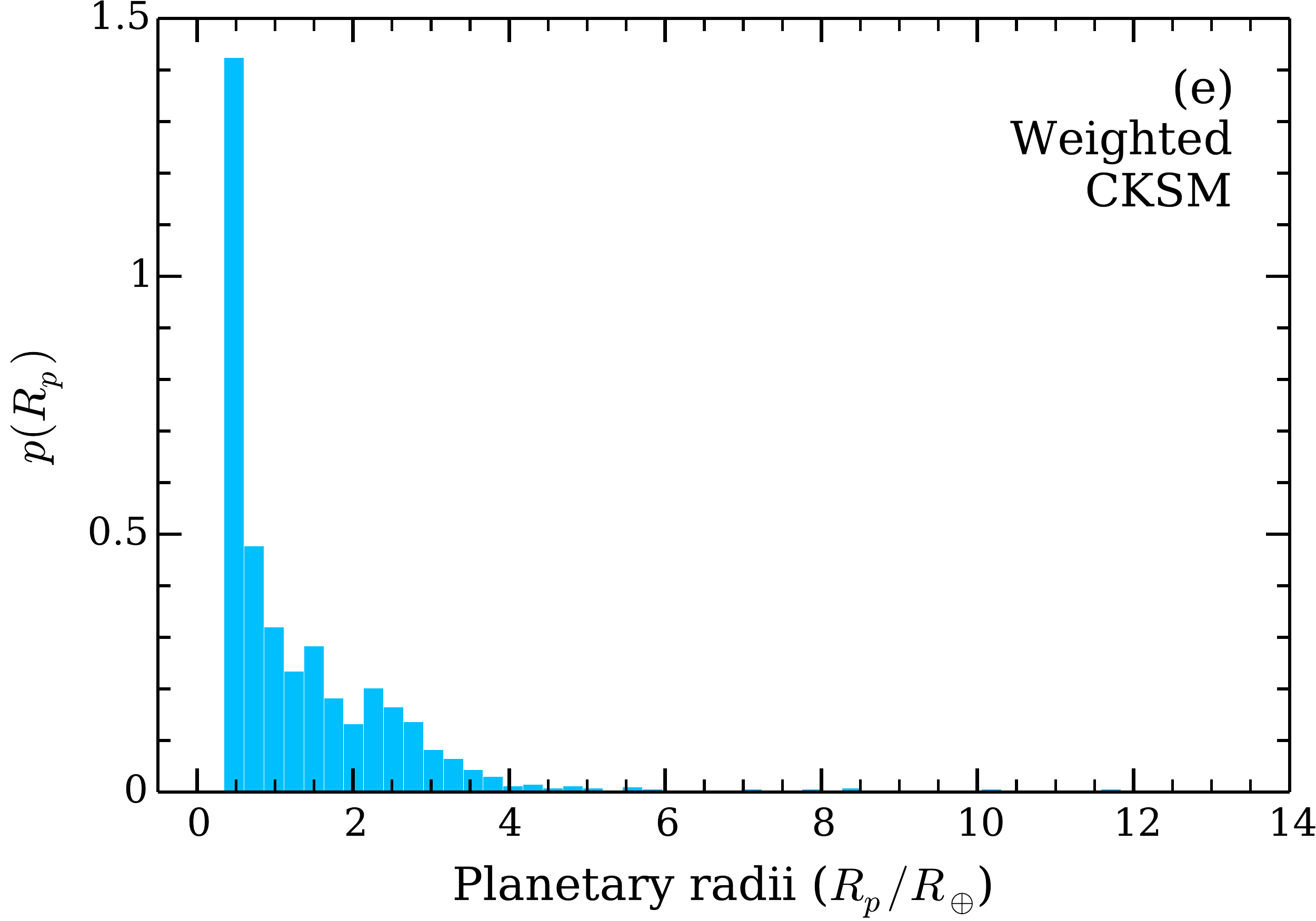}
\\
\includegraphics[width=0.45\columnwidth]{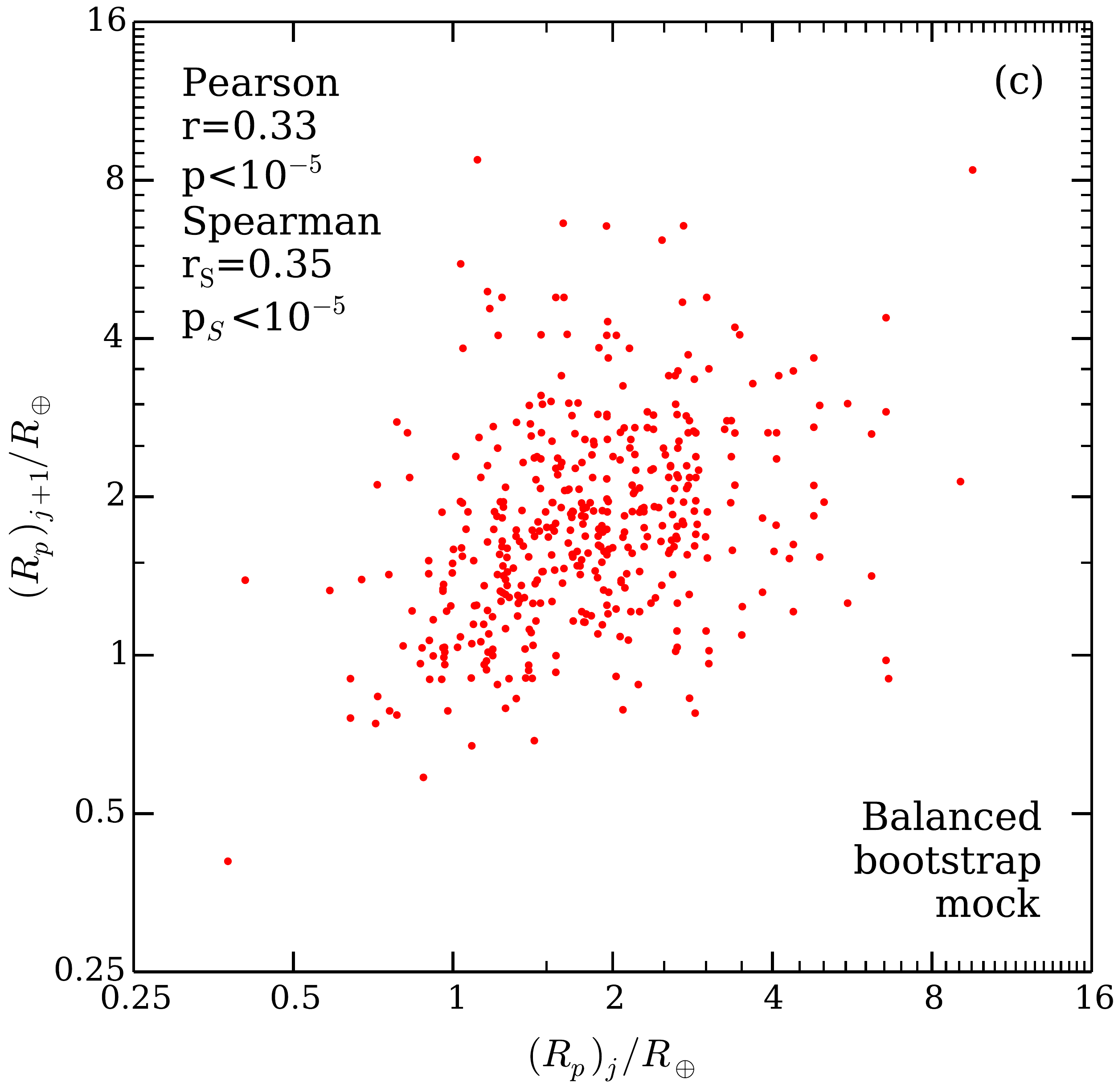} &
\includegraphics[width=0.45\columnwidth]{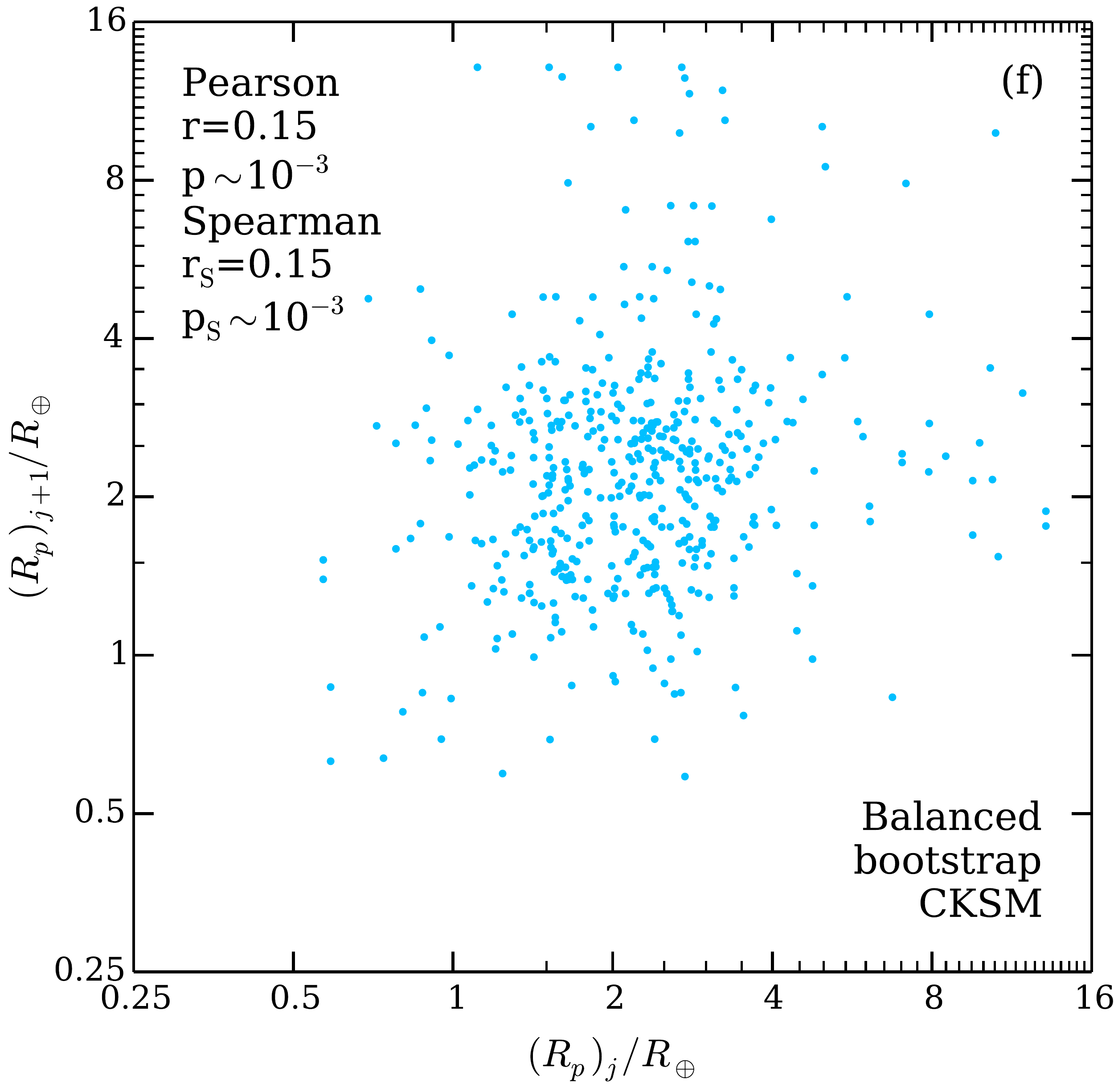}
\end{tabular}
\caption {Bootstrap tests. Red plots belong to the mock universe and blue to the observed planets. (a),(d) The results of a typical realization of the bootstrap test in which the planets are drawn from the observed distribution in Figure \ref{distributions}bd; these are directly comparable to the bootstrap tests done by W18. (b),(e) Distribution of planetary radii  weighted in detectability according to eq.\ (\ref{eq:w}).
(c),(f) The results of a typical realization of the balanced bootstrap simulation described in the text.  
In each scatter plot, the Pearson and Spearman correlation coefficients in $\log {R_p}$ and their corresponding $p$-value are printed in the upper left corner and represent the averages and medians of 5000 corresponding Monte-Carlo realizations.
The swapping criterion of W18 has been applied, i.e., a pair of planets is plotted only if the planets would still be detectable if their positions were swapped.}
\label{bootstrap}
\end{figure*}

To further examine the role of detection biases we repeat the bootstrap simulation in W18. We draw planetary radii at random with replacement from the observed radius distribution, mock or real, to populate the CKS positions, i.e. $\{\star_i,\mathrm{Per}_i\},$ then apply the SNR cut (\ref{eq:snr}). If the planet radius fails the cut, we draw another radius until all 909 available positions in 355 host systems are filled with detectable planets. The result of this approach is shown in Figure \ref{bootstrap}ad (Figure \ref{bootstrap}d is directly comparable to W18's Figure 4). In the mock universe we obtain a weak correlation with mean Pearson/Spearman $r= 0.19/0.21$ and median $p\sim 10^{-5}.$ In  the real universe we obtain Pearson/Spearman $r= 0.09/0.08$ with median $p= 0.11/0.06.$ These estimates are based on averaging 5000 Monte-Carlo trials in both the mock and real universes\footnote{We stress that the bootstrap algorithm involves somewhat different procedures for the real and mock universes. For the real universe we re-sample the original CKSM data set as there exists only one observational data set. In the mock universe each bootstrap realization is based on independently generating a new set of observed planets.}.

The failure of this bootstrap test to reproduce the strong correlation between the radii of adjacent planets found in the observed data (Figure \ref{distributions}e) led W18 to conclude that the correlation was likely driven by
astrophysics and not observational bias. This conclusion is premature as demonstrated by our mock universe: here the W18 bootstrap test is also unable to reproduce the strong observed correlation, but by construction the correlation is entirely due to observational bias. 

The bootstrap test used by W18 is flawed, because drawing the radii at random from the observed distribution gives preference to the most detectable planets over to the most abundant (for example, there are far fewer planets smaller than $1R_\oplus$ in Figure \ref{bootstrap}d than in Figure \ref{distributions}e). To remedy this flaw, 
we modified the W18 test by including a weighting procedure. After populating each position  $(\star_i,\mathrm{Per}_i)$ with an observable radius $R_{pj}$, we assign a weight $w_j$ to this radius proportional to the inverse of the number of positions at which it is detectable%
\footnote{Other weightings are possible, such as: 
(i) cycling $(R_{pj}, \Per_j)$ pairs through $(\star_i):$
$\tilde w_j^{-1} = \sum\limits_i \Theta[SNR(\star_i ; \Per_j, R_{pj})-10],$
where $i$ runs over all stars;
(ii) Cycling $R_{pj}$ through the shuffled set of $\Per_k$ and $(\star_i):$
$\tilde w_j^{-1} = \sum\limits_{i, k} \Theta[SNR(\star_i ; \Per_k, R_{pj})-10],$
where $i$ runs over all stars and $k$ over all periods;
and many others.
We present these weightings here to illustrate that the choice of weighting is non-trivial, and should be chosen to be consistent with the rest of the analysis and unambiguously stated.
\label{foot:two}
}: 
\be 
    w_j = \frac{\tilde w_j}{\sum \tilde w_k}, \quad \mbox{where} \quad 
    \tilde w_j^{-1} = \sum\limits_i \Theta \left[\mathrm{SNR}(\star_i ; \mathrm{Per}_i, R_{pj})-10\right],
    \label{eq:w}
\ee
where $\Theta[\cdot]$ is a step function. Then the empirical distribution $p(R_p)=\sum w_j\delta[R_p-R_{pj}]$ should be a more accurate estimate of the intrinsic distribution of radii. This approach is related to a technique known in the statistics literature as balanced bootstrap.

We first compare the distribution of radii derived in this way for the mock universe (Figure \ref{bootstrap}b) with the intrinsic distribution (Figure \ref{distributions}a). The intrinsic distribution is much steeper than the weighted distribution\footnote{
The following is a heuristic explanation of the problem. The smallest radius in the CKSM catalog is $0.34 R_\earth$, and a planet with this radius would be detectable in 5 of the 909 positions. In contrast, all 59 planets with $R_p > 4.47 R_\earth$ are detectable in all 909 positions (see Figure \ref{min_detectable}). As a result the detection probability of a planet with $R_p=0.34 R_\earth$ is only $909/5=182$ times smaller than the probability for planets with $R_p > 4.47 R_\earth .$ At the same time, in the intrinsic mock distribution the probability of occurrence per unit of log radius of $0.34 R_\earth$ planets is about 2,500 times higher than that of $4.47 R_\earth$ planets.}. We then used these weighted distributions to produce analogs of Figures \ref{distributions}ce, shown in Figures \ref{bootstrap}cf. We see that the correlations produced in the bootstrap analysis are stronger than in the W18 bootstrap analysis (compare Figure \ref{bootstrap}cf to \ref{bootstrap}ac) but still substantially weaker than in the original data, mock or real (Figures \ref{distributions}ce). Averaged over 5000 realizations, balanced bootstrap produces Pearson/Spearman $r=0.15$ and median $p=0.001$ in the CKSM data, and Pearson/Spearman $r=0.33/0.35$ and $p\ll 10^{-5}$ in the mock data, compared to Pearson/Spearman $r\sim 0.6$ in the original data for both the real and the mock universes.

We conclude that neither the W18 bootstrap analysis nor the balanced bootstrap described by equation (\ref{eq:w}) offers a correct description of the statistics of either the mock or the real data in the CKSM catalog. 
Informally, bootstrap fails because the data are truncated---planets that lie below some minimum radius are not included in the catalog, and this cutoff varies from star to star. 

We can illustrate this problem with a simple example. Consider a population of $N\gg1$ stars that has two types of planet, one of radius $R_1$ (``small" planets) and the other of radius $R_2>R_1$ (``large" planets). The probability that a given star will host a small or large planet is $p_1$ or $p_2$, and these two events are independent; we also assume for simplicity that $p_1$ and $p_2$ are small enough that the fraction of stars containing more than one planet is negligible. Finally, small planets can be detected in a fraction $f$ of the stars while large planets can be detected in all the stars. With these assumptions, the intrinsic ratio of the number of small planets to the number of large planets is $r=p_1/p_2$. 

We now compile a catalog of planets from this population. There are $Np_1$ small planets of which a fraction $f$ are detectable, so the catalog will contain $N_1=Np_1f$ small planets. There are $Np_2$ large planets, which are all detectable, so the catalog contains $N_2=Np_2$ large planets; of these, $N'_2=Np_2f$ are in systems where a small planet could be detected. The complete catalog contains $N_\mathrm{cat}=N_1+N_2=N(p_1f+p_2)$ planets or stars. The fraction of these in which a small planet could be detected is $(N_1+N_2')/N_\mathrm{cat}$ which we set equal to an inverse weight, $w^{-1}$. Then according to the balanced bootstrap procedure based on equation (\ref{eq:w}), we estimate that the ratio of the number of small planets to the number of large planets is
\begin{equation}
    r_\mathrm{est}=\frac{N_1w}{N_2}=r\frac{p_1f+p_2}{p_1+p_2},
    \label{eq:app.r}
\end{equation}
which is less than the true ratio $r$ unless $f=1$.

In the CKSM sample we have 909 planets, out of which 59 are large enough to be detectable at every position -- these are the ``large" planets in our example.
%\footnote{\lena{Alternatively, one can eliminate K00881 system, the one with $R_\max=4.47 R_\earth,$ entirely and consider 907 planets in 354 systems with the largest minimal detectable radius of $3.1 R_\earth.$ Then we will have 154 large planets and 752 small ones. This would not, however, change our conclusion. Note, that $3.1 R_\earth$ is almost exactly the same radius at which the correlation between $R_{pj}$ and $R_{pj+1}$ in Figure \ref{distributions}e (see Figure \ref{pearsons}a), becomes unreliable. \scott{I'm not sure I like this footnote...I think it's too much detail...}}}. 
The remaining 850 are ``small" planets. This implies that in our example $p_2 \ll p_1$ or $r \gg 1$ and thus in doing a bootstrap analysis similar to the one in this Section one would arrive at the false conclusion that the intrinsic ratio of small to large planets is $r_\mathrm{est}\simeq r f+1$---the limit of equation (\ref{eq:app.r}) if $r\to\infty$ while $rf$ remains constant (if no detectability weighting is used at all $r_\mathrm{est}=r f$).
Thus the balanced bootstrap simulations underestimate the occurrence rate of small planets by the fraction of positions in which they are detectable, which can be smaller than unity by a few orders of magnitude.%percent. 

An important feature to note about the CKSM catalog is that it contains only systems with detected planets. One might ask whether the balanced bootstrap procedure of equation (\ref{eq:w}) would work if applied to a larger catalog, compiled with uniform selection criteria, in which only \emph{some} of the systems had detected planets, i.e., the catalog before the stars with no detected planets are removed. This is the approach used for example by \citet{2013PNAS..11019273P} and \citet{2017AJ....154..109F}. However this procedure only works under special conditions, a key one of which is universality of all observed planetary systems and planet formation properties therein, see the Appendix.

% 5000 CKS [1=observed,2=Weiss boot,3=weighted boot]:
%Pearsons
%>>> print mean(rp1), median(pp1)
%0.650240449536 2.23136111545e-62
%>>> print mean(rp2), median(pp2)
%0.070757714149 0.112444281202
%>>> print mean(rp3), median(pp3)
%0.14764144747 0.00104913494799
%>>> print "Spearmans"
%Spearmans
%>>> print mean(rs1), median(ps1)
%0.573899414966 7.77739474954e-46
%>>> print mean(rs2), median(ps2)
%0.0838638086538 0.0564028317927
%>>> print mean(rs3), median(ps3)
%0.148404606939 0.000958128420473

% 5000 trials in mock universe [1=observed,2=Weiss boot,3=weighted boot]:
%>>> print mean(rp1), median(pp1)
%0.562360692315 7.37705409742e-32
%>>> print mean(rp2), median(pp2)
%0.192244856055 2.90256759038e-05
%>>> print mean(rp3), median(pp3)
%0.332577655752 1.07048795952e-12
%>>> print "Spearmans"
%Spearmans
%>>> print mean(rs1), median(ps1)
%0.591134349755 9.29352704198e-36
%>>> print mean(rs2), median(ps2)
%0.214812993352 2.70874660524e-06
%>>> print mean(rs3), median(ps3)
%0.353191107555 3.46218191069e-14

%%%%%%%%%%%%%%%%

\begin{figure}
\vspace{-0.0cm}
\centering
\begin{tabular}{c}
\includegraphics[width=0.48\columnwidth]{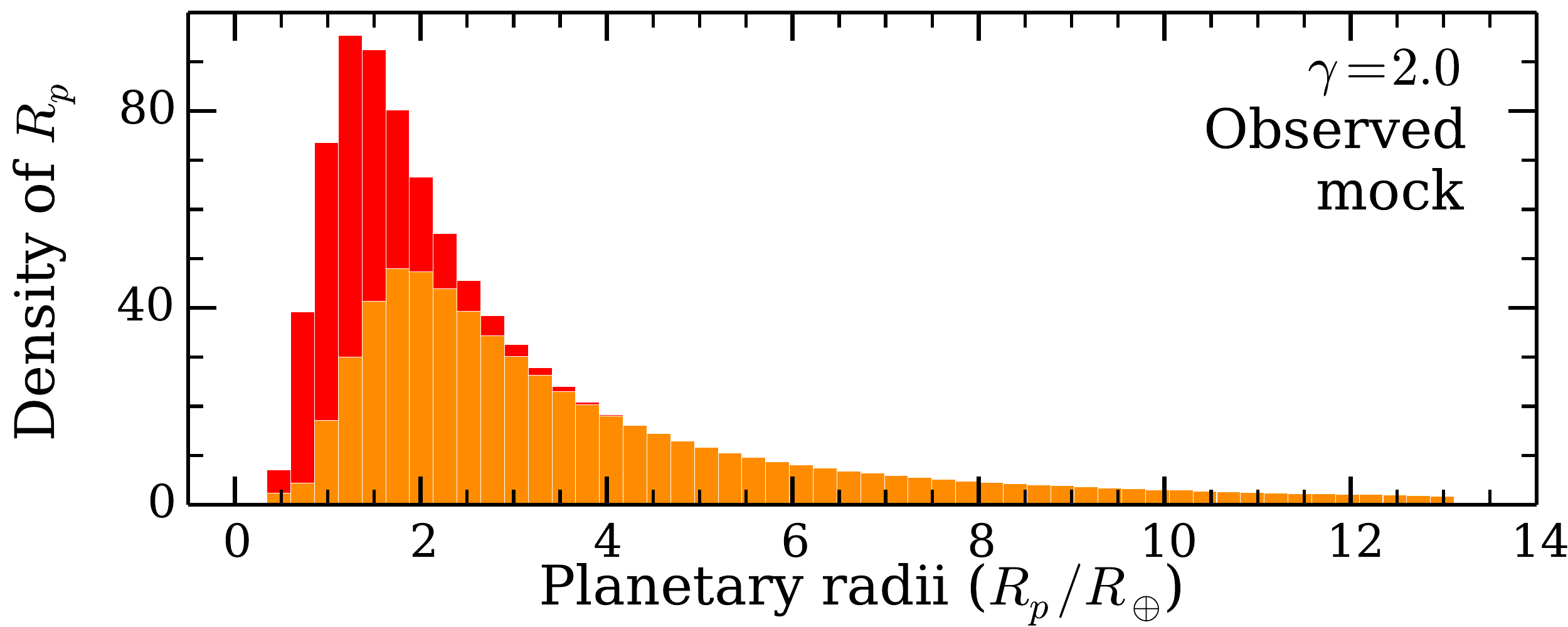}\\
\includegraphics[width=0.48\columnwidth]{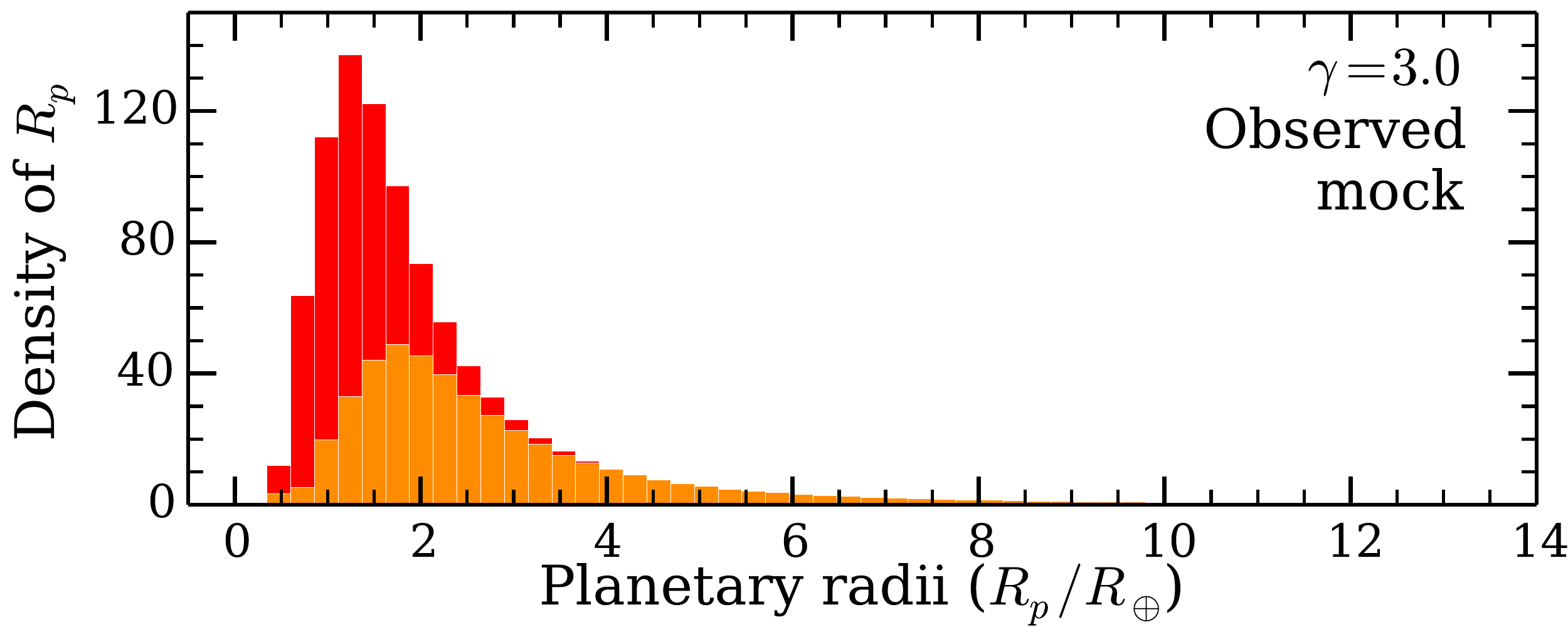}\\
\includegraphics[width=0.48\columnwidth]{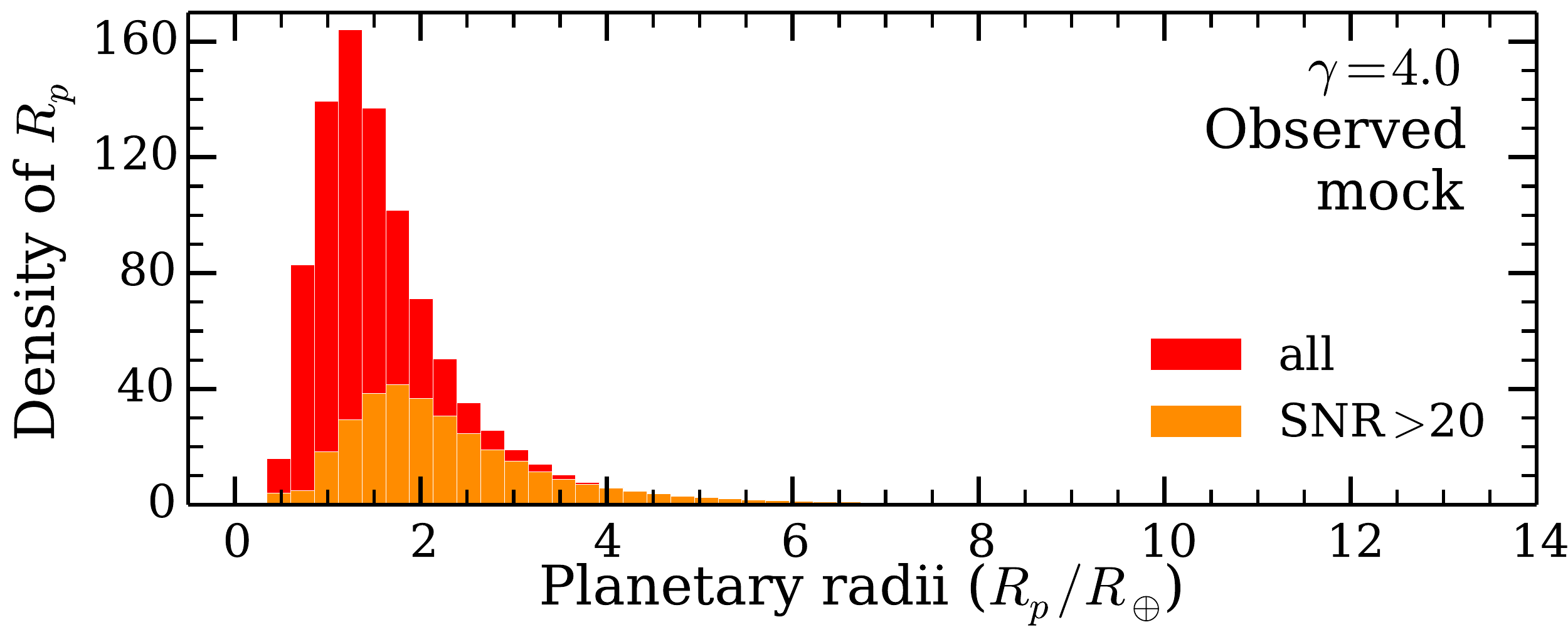}
\end{tabular}
\caption {The mock observed planetary radii distribution (red) as a function of the exponent $\gamma$ of the power-law distribution (eq. \ref{eq:gamma}). The distributions of the planets with SNR$>20$ are shown in orange. The plots each show an average of 5000 Monte-Carlo realizations. For the corresponding distributions in the CKSM catalog see Figure \ref{fig:snr.fit}c.}
\label{fig.gamma.snr}
\end{figure}

\begin{figure}
\vspace{-0.0cm}
\centering
\includegraphics[width=0.48\columnwidth]{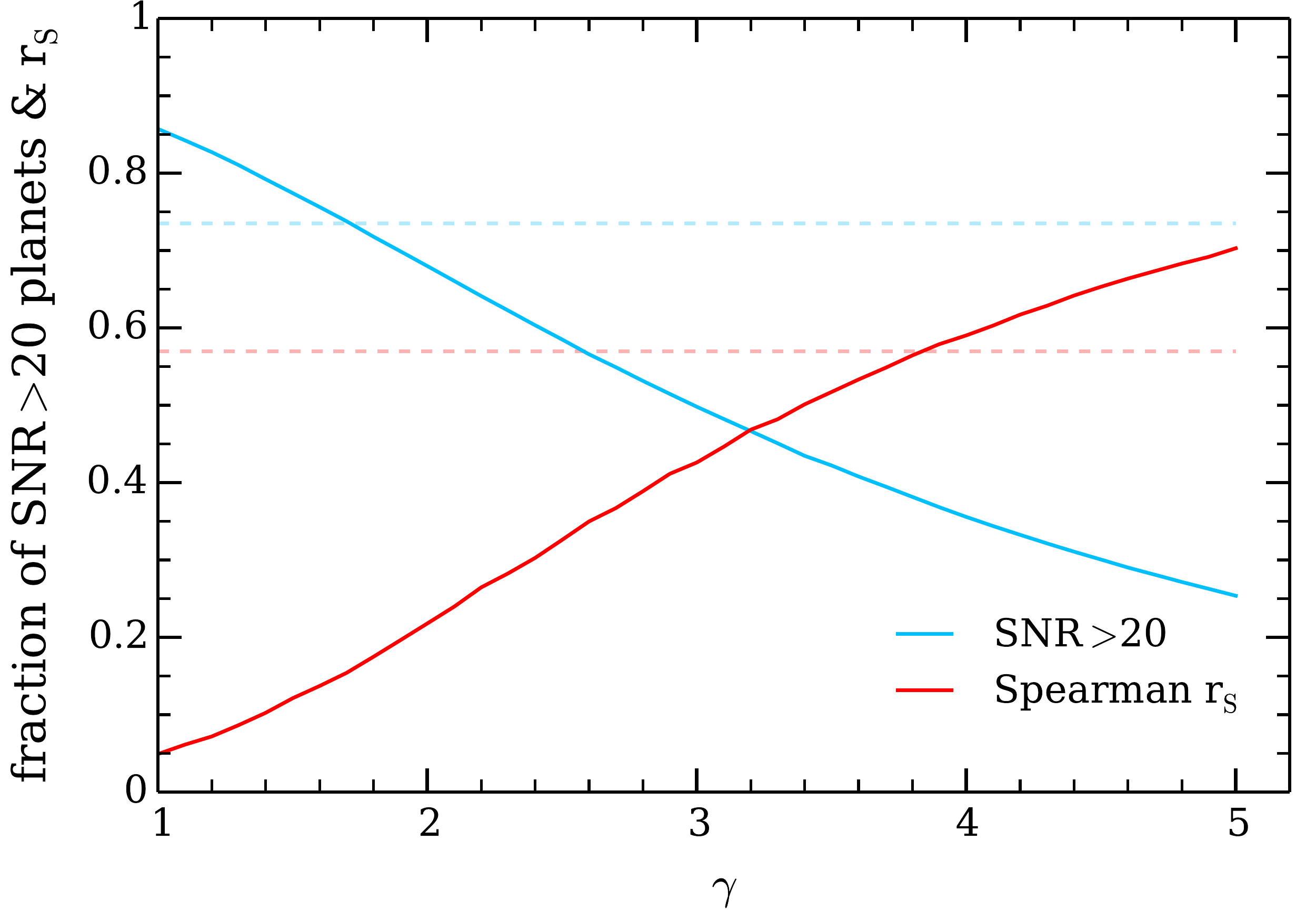}
\caption {The value of the Spearman correlation coefficient $r_\mathrm{S}$ between the radii of adjacent planets (red) and the fraction of the planets with SNR$>20$ in the mock observed planets (blue) as a function of he exponent $\gamma$ of the power-law distribution (eq.\ \ref{eq:gamma}). The faint dashed lines of the corresponding colors represent the observed values from the CKSM catalog.}
\label{fig.gamma.var}
\end{figure}

\bigskip

\section{On the distribution of planetary radii}\label{sec:gamma}

In our mock universe we set the intrinsic distribution of radii to a power law (eq.\ \ref{eq:gamma}) with exponent $\gamma=4$. Larger values of $\gamma$ lead to a tighter correlation between $(R_{p})_{j}$ and $(R_{p})_{j+1},$ but produce fewer large planets ($R_p\gtrsim 4R_\oplus$), see Figures \ref{fig.gamma.snr} and \ref{fig.gamma.var}. We chose the simple integer exponent $\gamma = 4$ because it produces a correlation between the radii of the adjacent planets that is close to the observed Spearman $r_\mathrm{S} = 0.57,$ a sizable population of large planets, and a scatter plot of $(R_{pj}, R_{p,j+1})$ points similar to the one observed in the CKSM catalog.

Other authors have carried out much more careful modeling of the planetary radius distribution. \cite{2015ApJ...809....8B} fit the distribution of radii between $0.7R_\oplus$ and $2.5R_\oplus$ to single and broken power laws. For single power laws they find $\gamma=1.5$ with an allowed range from $-0.5$ to $3.25$, although models with broken power laws are preferred. \citet{2013PNAS..11019273P} find that the distribution is nearly flat in logarithmic radius (i.e., $\gamma=-1$) at the smallest radial range detected (1--$2R_\oplus$). A more recent study by
\cite{2017AJ....154..109F}, which employs the CKS data on single-planet systems, only considers planets with $R_p>1.14 R_\earth$ and finds a significant gap in the radius distribution between $1.3R_\oplus$ and $2.4R_\oplus$. \cite{2017ApJ...849L..33M} fit the distribution to a broken power law with $\gamma=-0.5\pm0.2$ for radii less than $3R_\oplus$. \cite{2019AJ....158..109H} estimate the radius distribution between $0.5R_\oplus$ and $R_\oplus$ and their results can be fit approximately by $\gamma\approx 1$. Our distribution is steeper than any of these, but it is also important to point out that for the planets with $R_p \lesssim R_\earth$ our knowledge of the occurrence rate is quite limited. Despite these limitations, our toy model produces a distribution of observed planets that is similar to the distribution in the real CKSM catalog (compare panels b and d in Figure \ref{distributions}). This toy model is a sufficient tool to illustrate the failure of the W18 bootstrap in de-biasing the observed radius distribution.

An obvious flaw of the simple power-law model is overproduction of small planets, which in turn implies the inability to recover the SNR distribution:  in the CKSM data the fraction of planets with SNR$>20$ is 74\%, while in our mock universe on average only about 35\% of the planets have SNR$>20$ (see Figures \ref{fig.gamma.snr} and \ref{fig.gamma.var}). We address this issue in Section \ref{sec.snr.fit}.

\begin{figure}
\vspace{-0.0cm}
\centering
\includegraphics[width=0.48\columnwidth]{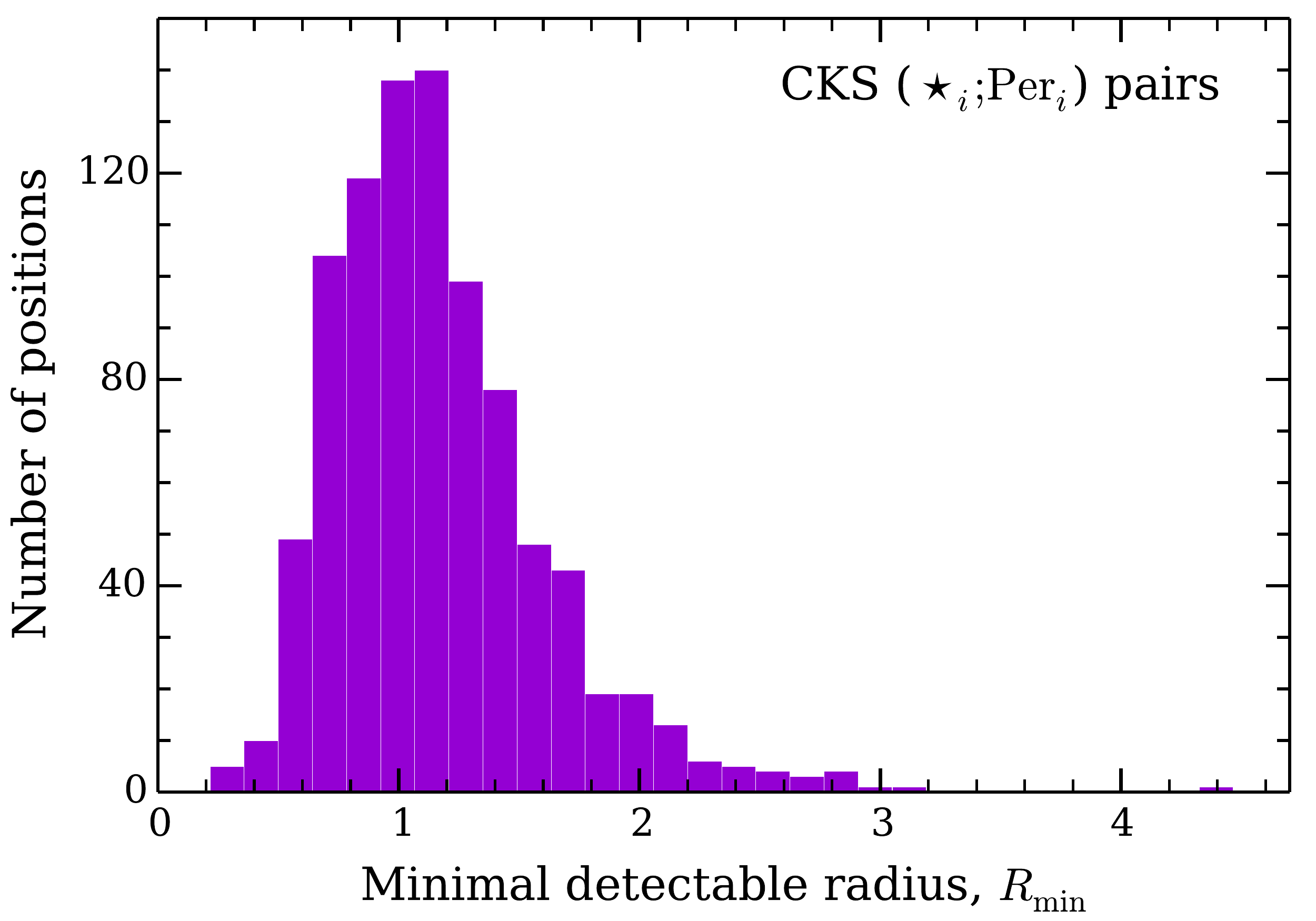}
\caption {Smallest detectable radii. The number of available positions $(\star_i,\mathrm{Per}_i)$ as a function of the minimal detectable radius according to equation (\ref{eq:snr}).}
\label{min_detectable}
\end{figure}

\begin{figure*}
\vspace{-0.0cm}
\centering
\begin{tabular}{cc}
\includegraphics[width=0.48\columnwidth]{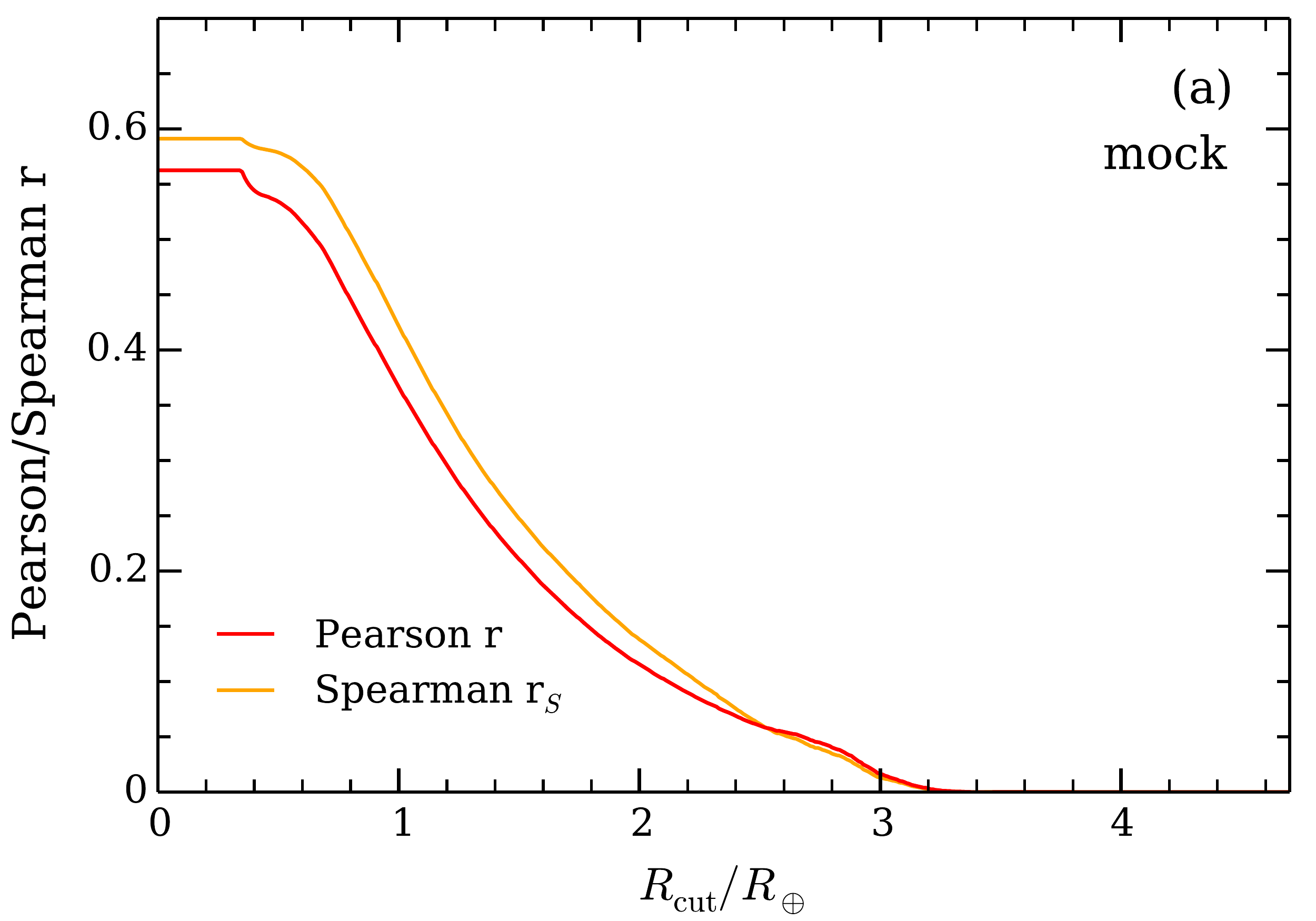} &
\includegraphics[width=0.48\columnwidth]{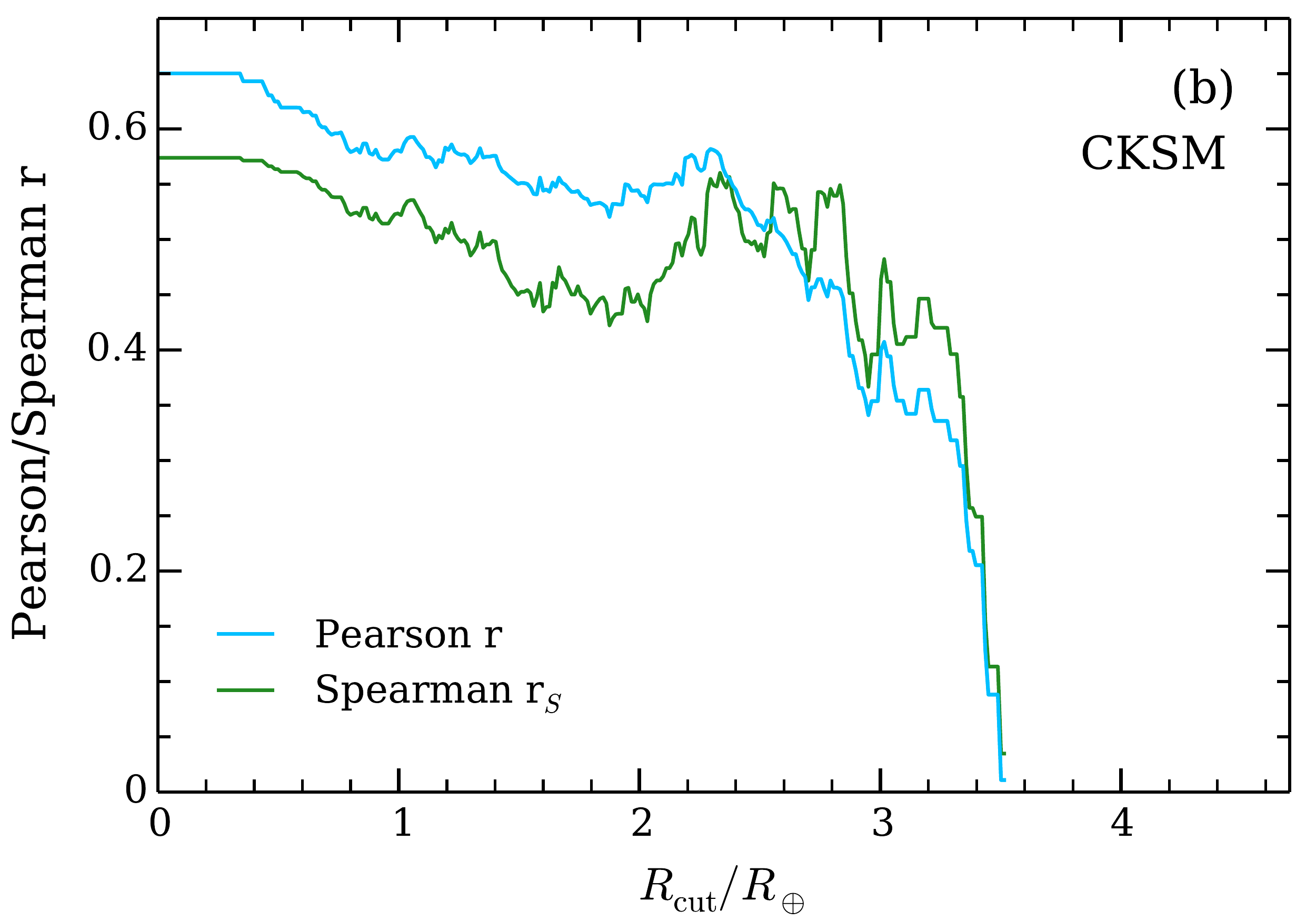}
\end{tabular}
\caption {Pearson and Spearman correlation coefficients as a function of the minimum radius $R_\mathrm{cut}$ of planets included in the correlation, i.e., $R_{pj}, R_{p,j+1} \geq R_\mathrm{cut}.$ (a) Pearson $r$ (red) and Spearman $r_s$ (yellow) for the mock universe. We plot the average of 25,000 Monte-Carlo realizations.  The median Pearson/Spearman $p$-value is $<0.01$ for $R_\mathrm{cut}\le 1.47/1.56R_\oplus$ and rises to $> 0.4$ by $R_\mathrm{cut}\simeq 2.2 R_\oplus.$ (b) Pearson $r$ (blue) and Spearman $r_s$ (green) for CKSM data. The Pearson/Spearman $p < 0.03/0.01$ at $R_\mathrm{cut} \leq 3.3 R_\earth$, showing that the correlation is significant at a level of a few percent or better, but rises steeply to $>0.5/0.6$ at $R_\mathrm{cut} = 3.45 R_\earth.$
%
%\lena{Combining the 25,000 Monte-Carlo realizations into one data set allows is to trace the correlation to the point of its vanishing, which happens at $\sim 4.1/4.2 R_\earth$ when Pearson/Spearman $r \sim 0.01.$}
}
\label{fig.pearsons}
\end{figure*}

\bigskip

\section{Correlations of radii of large planets}

\label{sec:large}

W18 comment that there are many pairs of large planets with both radii close to $8R_\oplus$, and that this correlation cannot arise from observational selection effects because all planets with these large radii are detectable. 
%A close visual inspection of Figures \ref{distributions}c and \ref{distributions}e confirms the lack of planets with $R_p\gtrsim 4R_\oplus$ in the mock universe (red) compared to %the correlation looks stronger in 
%the real data (blue). 
But is the correlation between the radii of adjacent planets significant for large radii? To quantify this, we have cut off the observed distributions in both the mock universe and the CKSM sample at a minimum radius $R_\mathrm{cut}$ and plotted the correlation coefficients as a function of $R_\mathrm{cut}$ in Figure \ref{fig.pearsons}. We see that no significant correlation between the radii of adjacent planets with $R_{p} > 3.4R_\oplus$ is present in the CKSM data. 

Nevertheless there are differences between the power-law toy model and the data. In particular, the correlation between $R_{pj}$ and $R_{p,j+1}$ in the CKSM sample persists up to a cutoff radius $R_\mathrm{cut}\simeq 3.4R_\oplus$, whereas the correlation declines in the mock catalogs for $R_\mathrm{cut}\gtrsim 1 R_\oplus$ and becomes weak by about $1.5 R_\earth$ (see left panel of Figure \ref{fig.pearsons}). This is not surprising as the correlation in the mock universe is due to observational biases, which are dominated by the distribution of smallest detectable radii (Figure \ref{min_detectable}). This distribution is rapidly falling  for $R_p \gtrsim 1.2 R_\earth.$ If the width of the distribution of the minimal detectable radii were larger, the correlation should persist to larger radii.

We conclude that the extremely simple model of hypothesis (i)---``planets don't know anything"---can successfully reproduce the overall correlation between adjacent planet radii seen in W18 (Figure \ref{distributions}). In this sense it provides a counter-example to W18's arguments that the observed correlation must be astrophysical. However, other data indicate that hypothesis (i) cannot be the whole explanation: the SNR distribution is not consistent with the data, the correlation is weaker than observed for radii between $R_\oplus$ and $3R_\oplus$, and the radius distribution is too steep. We therefore turn to hypothesis (ii).

%These observations confirm the visual impression that the mock universe has difficulty reproducing the radius correlations among planets in the limited range 1--$3R_\oplus$. We do not regard this as a major weakness of our model as anomalies of this kind can be present in any model if one looks closely enough.

\bigskip

\section{``Planets know about the system they formed in''. Reproducing the SNR distribution.}\label{sec.snr.fit}

In this Section we complicate the rules guarding our mock universe. We assume that ``planets know about the system they formed in'', that is, that the distribution of planetary radii may vary from system to system. This is a very plausible hypothesis, since the distribution of radii presumably depends on the properties of the protoplanetary disk which differ for every star. 
We demonstrate that with this assumption we can reproduce the distribution of planetary radii observed in the CKSM catalog, the distribution of SNR, and even the correlation coefficients between the radii of adjacent planets.

We consider four types of planetary systems ($\a_a,\, \a_b, \, \a_c, \, \a_d$):

\begin{enumerate}[(a)]
    \item Systems with predominantly small planets. For these systems, similarly to Section \ref{sec:creation}, we set the planet radius distribution to a power-law $p(R_p|\a_a) \sim R_p^{-\gamma};$

    \item Systems in which planets tend to have $R_p \simeq 1.3 R_\earth$. We model the radius distribution as a log Gaussian,
    \begin{equation}
        p(R_p|\a_b) \sim \exp \left[ - \frac{(\log (R_p)-\log(R_1))^2}{2 \sigma_1^2} \right];
    \end{equation}

    \item Systems in which planets tend to have $R_p \simeq 2.4 R_\earth$. We again model the radius distribution with a log Gaussian,
    \begin{equation}
        p(R_p|\a_c) \sim \exp \left[ - \frac{(\log (R_p)-\log(R_2))^2}{2 \sigma_2^2} \right];
    \end{equation}

    \item Systems with large planets. Here $p(R_p|\a_d) \sim R_p^{-1}$ for $R_p \geq R_d$ and $p(R_p|\a_d)=0$ for $R_p< R_d.$ All such large planets are detectable and thus one may expect that their observed distribution is close to their intrinsic distribution. 
    
\end{enumerate}

To generate a set of mock CKSM planetary systems, we randomly assigned CKSM systems to one of the above distributions with probability $\Prob_\mu,$ where $\mu = a,b,c,d$ and $\sum_{\mu} \Prob_\mu =1,$ using a similar procedure to the one discussed in Section \ref{sec:creation}\footnote{In the case of a randomly assigned distribution failing to deliver detectable planets in a given system, which we defined as 10,000 failed attempts, we reassigned the system to another mock distribution. The reassignments are rare, happening on average for 1\% of the systems.}. For each set of nine parameters $\gamma,$ $R_1,$ $\sigma_1,$ $R_2,$ $\sigma_2,$ $R_d,$ $\Prob_b,$ $\Prob_c$ and $\Prob_d$ we generate 100 mock systems and average their parameters.
We evaluate the quality of the fit between the mock systems and the CKSM catalog based on the following criteria: (1) Maximizing the p-value of the Kolmogorov--Smirnov (KS) comparison test between the mock and true observed planetary radius distributions; (2) Maximizing the p-value of the KS comparison between the mock and true observed planetary radius distributions detected with SNR$>20$; (3) the Spearman correlation coefficient $r_p$ of the correlations between adjacent planetary radii must be close to the observed value; (4) The fraction of $R_{p,j+1} (R_{p,j})$ pairs that fall into the boxes $(R_{p,j+1}, R_{p,j}) < 1 R_\earth,$ $1 R_\earth < (R_{p,j+1}, R_{p,j}) < 2 R_\earth,$ $2 R_\earth < (R_{p,j+1}, R_{p,j}) < 4 R_\earth,$ and $(R_{p,j+1}, R_{p,j}) > 4 R_\earth$ must be close to the values observed in the CKSM catalog\footnote{For consistency of the comparison tests we removed the system K03158. All five planets in the system were obvious outliers in the SNR($R_p$) distribution.}\label{foot:drop}.

\begin{figure*}
\vspace{-0.0cm}
\centering
\begin{tabular}{cc}
\includegraphics[width=0.45\columnwidth]{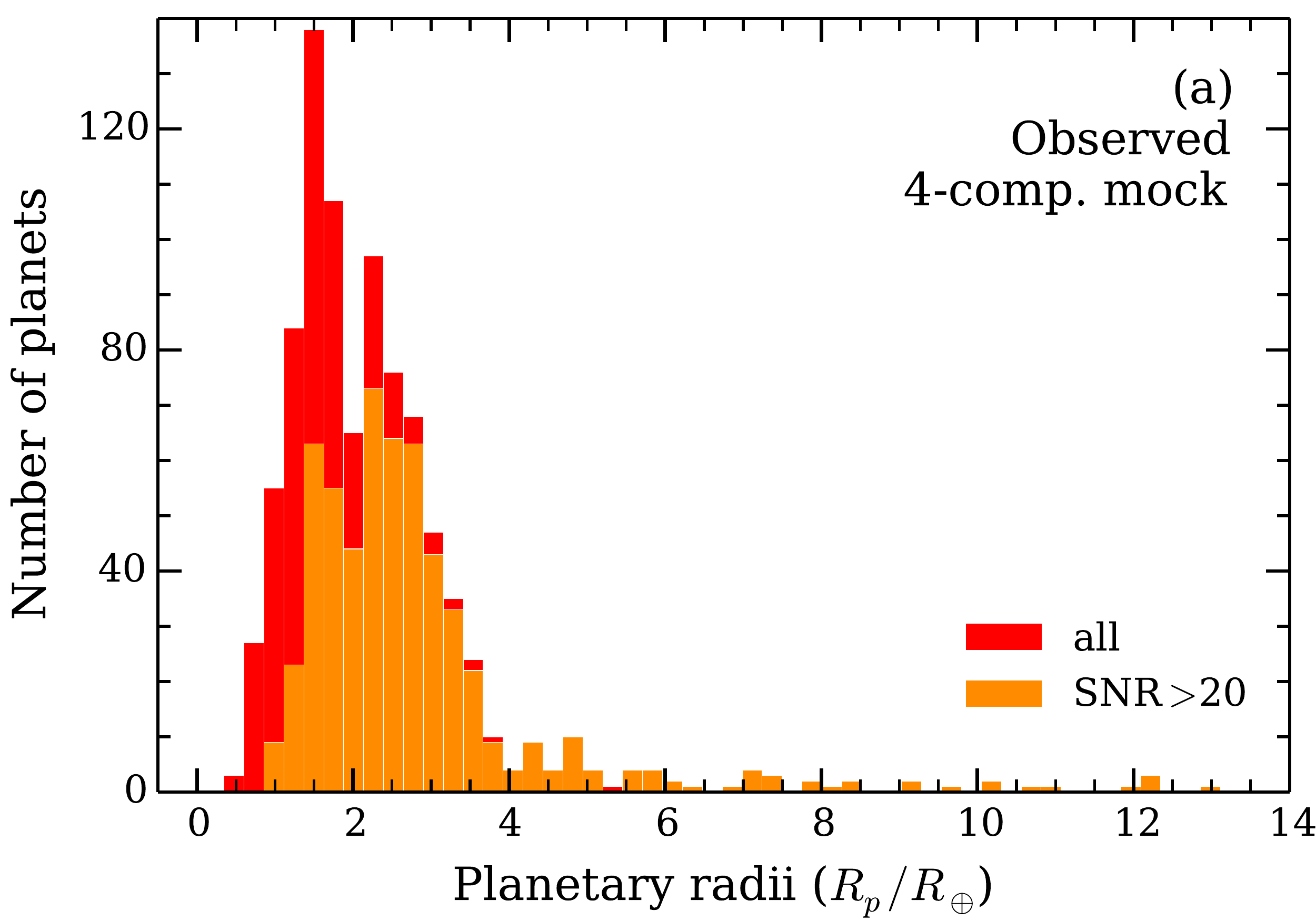} &
\includegraphics[width=0.45\columnwidth]{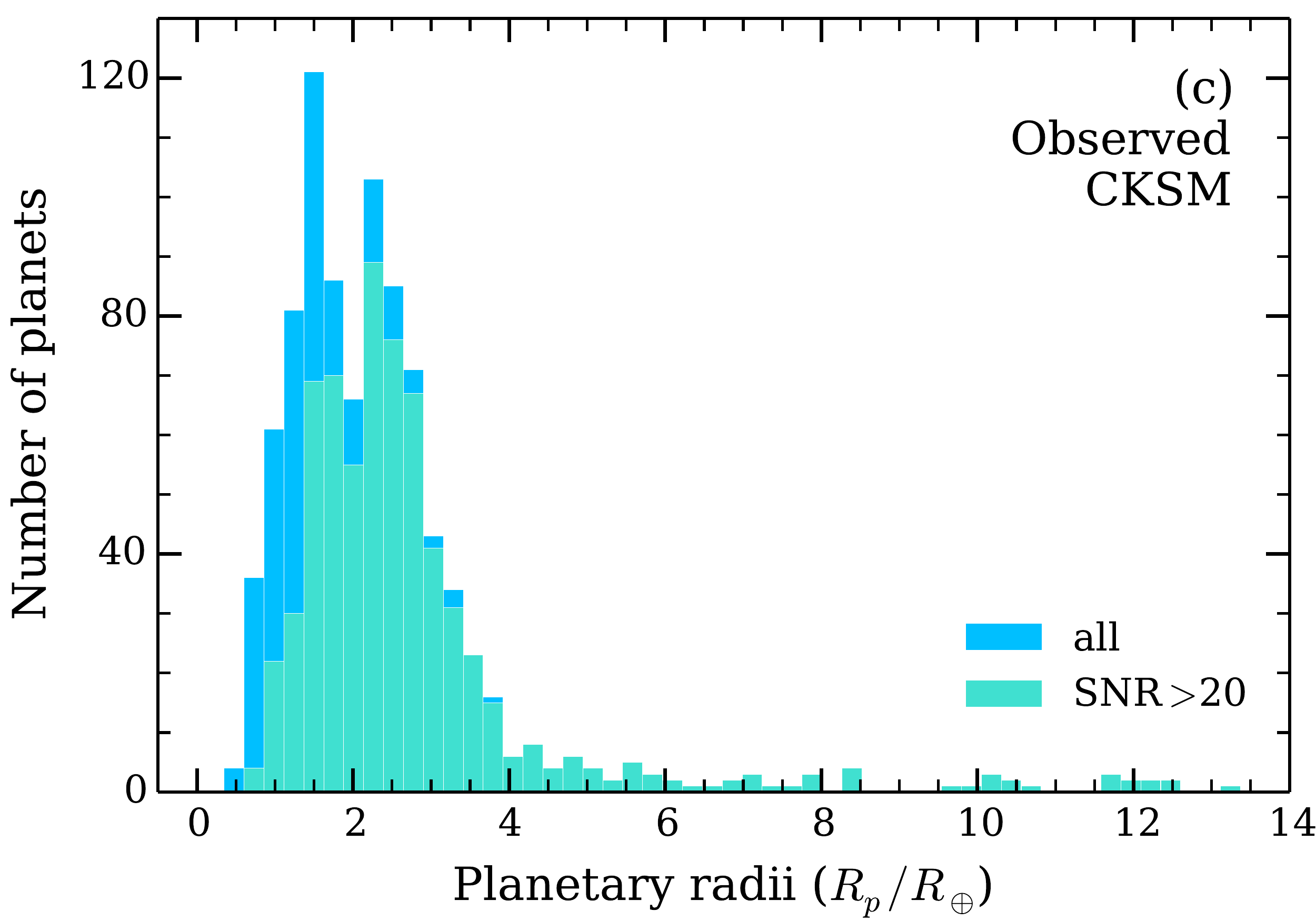}
\\
\includegraphics[width=0.45\columnwidth]{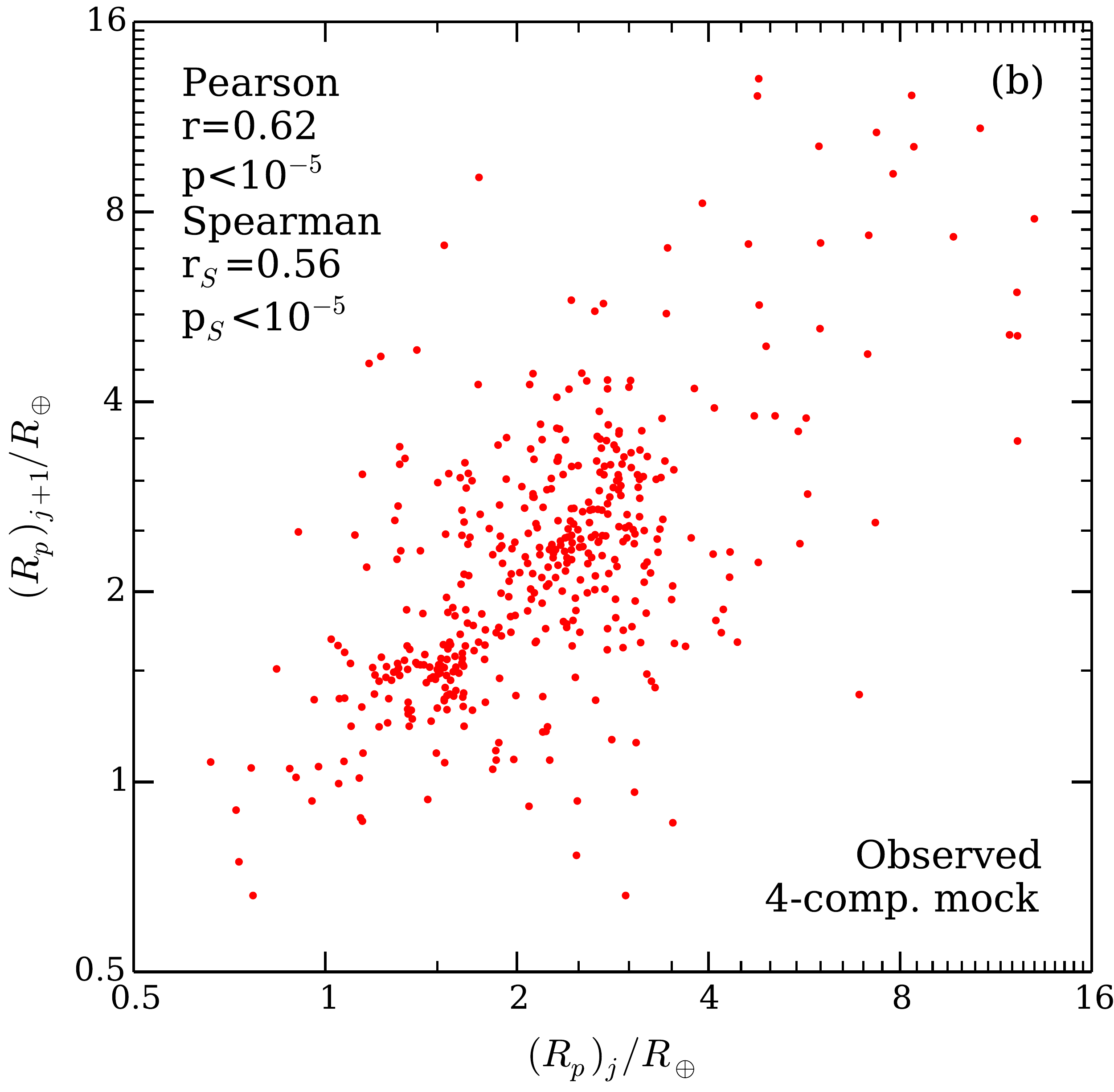} &
\includegraphics[width=0.45\columnwidth]{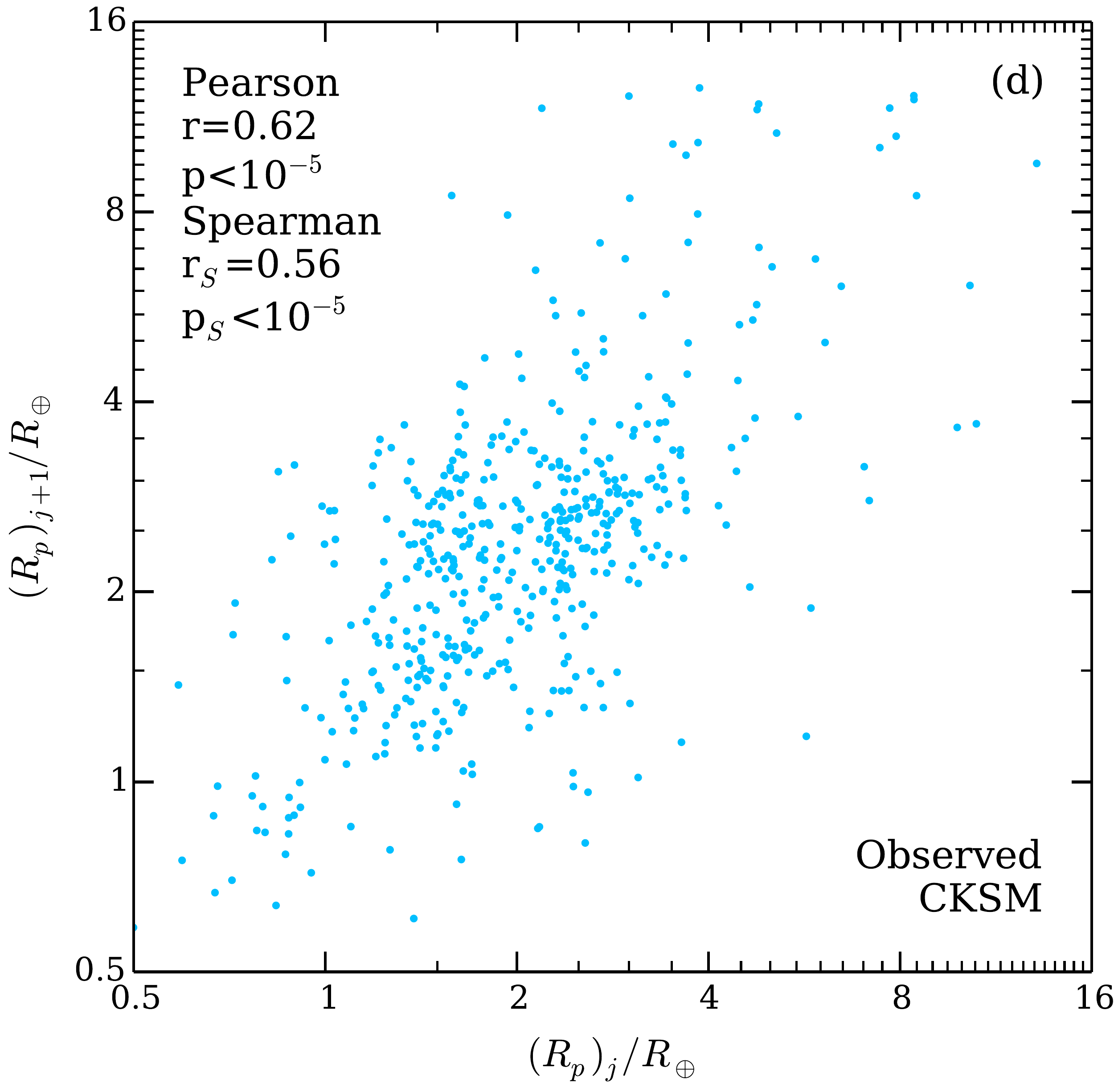}
\end{tabular}
\caption {Observations in the four-component mock universe compared to the CKSM sample. Red/orange plots belong to an example Monte-Carlo realization in the mock universe and blue/green to the observed planets. (a) The distribution of radii of the observed mock planets. We randomly assign each of the 354 CKSM systems (see footnote \ref{foot:drop}) to one of the four distributions described in Section \ref{sec.snr.fit} and for each available $(\star_i, \mathrm{Per}_j)$ position in the system, we draw a radius at random from the assigned mock distribution, apply the SNR cut of equation (\ref{eq:snr}), and repeat until the planet survives the cut. (b) The relation between the radii of adjacent planets $R_{pj}$ and $R_{p,j+1}$ for the Monte-Carlo realization of mock planets plotted in (a). 
(c) and (d) The distribution of radii and the relation between the radii of adjacent planets in the CKSM sample. Concerning the difference with Figure \ref{distributions}e refer to footnote \ref{foot:drop}. In panels (b) and (d) we applied the swapping criterion of W18, i.e., a pair of adjacent planets is plotted only if the planets would still be detectable if their positions were swapped.}
\label{fig:snr.fit}
\end{figure*}

\begin{figure}
\vspace{-0.0cm}
\centering
\includegraphics[width=0.48\columnwidth]{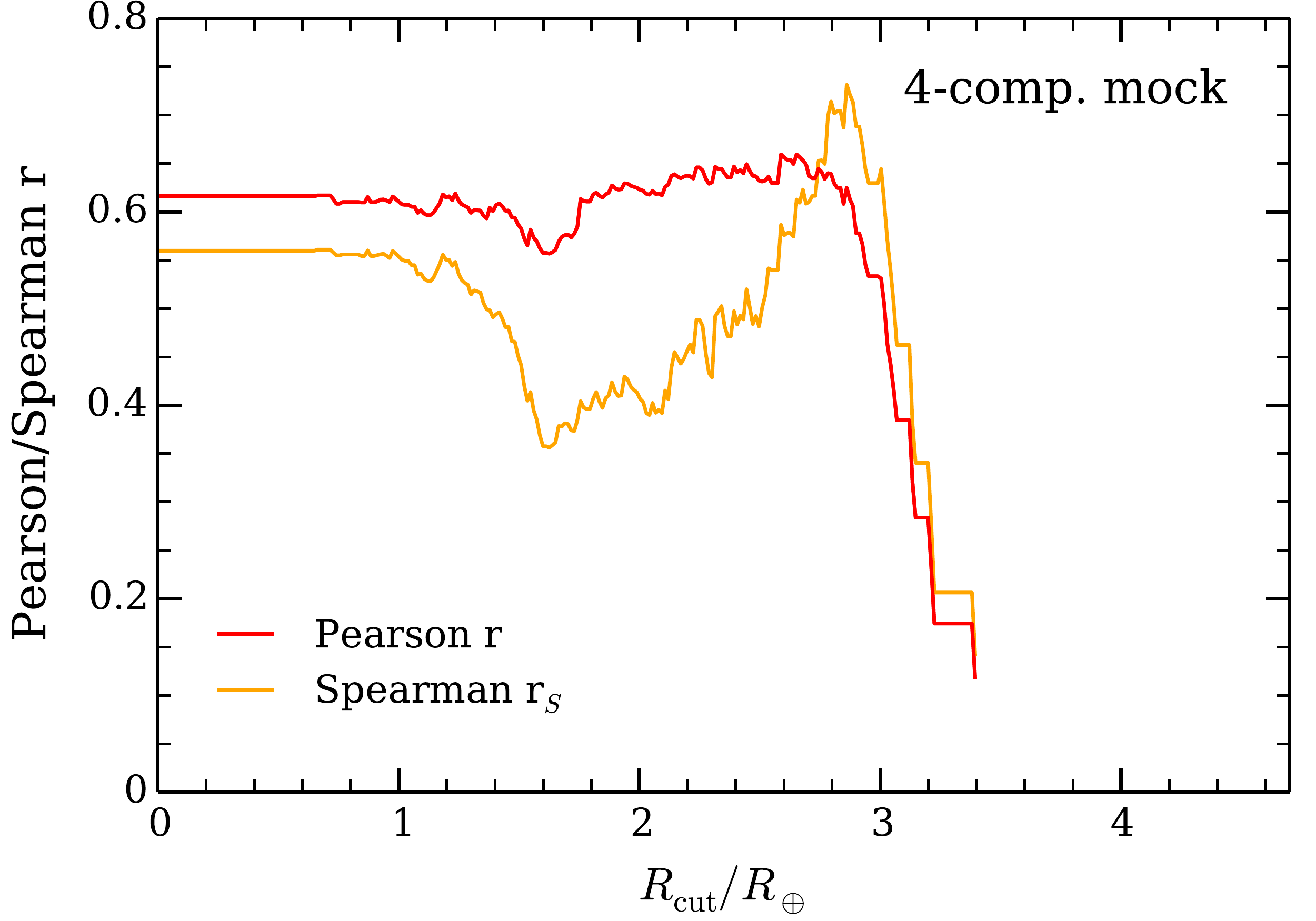}
\caption {Pearson and Spearman correlation coefficients as a function of the minimum radius $R_\mathrm{cut}$ of planets included in the correlation, i.e., $R_{pj}, R_{p,j+1} \geq R_\mathrm{cut},$ for the four-component Monte-Carlo realization in Figure \ref{fig:snr.fit}ab. The Pearson/Spearman $p < 0.02/0.01$ at $R_\mathrm{cut} \leq 3.1 R_\earth$, showing that the correlation is significant at a level of a few percent or better, but rises steeply to $\sim0.3$ at $R_\mathrm{cut} = 3.2 R_\earth$ and to $\sim0.5$ at $R_\mathrm{cut} = 3.4 R_\earth.$ Compare Figure \ref{fig.pearsons}.}
\label{fig:snr_spearmans}
\end{figure}

Figure \ref{fig:snr.fit} presents an example Monte-Carlo realization of the four-component universe side-to-side with the CKSM data. The parameters used to generate this realization are
\be
\ba{c}
    \displaystyle \g=3.2, \quad R_1 = 1.42 R_\earth, \quad \sigma_1 = 0.12, \quad R_2 = 2.35 R_\earth, \quad \s_2 = 0.23, \quad R_d = 3.4 R_\earth \\ 
    \displaystyle \Prob_a=1-\Prob_b - \Prob_c -\Prob_d, \quad \Prob_b+\Prob_c = 0.54, \quad  {\displaystyle \Prob_b}/{(\displaystyle \Prob_b+\Prob_c)}=0.3, \quad \Prob_d = 0.035.
\ea\label{eq:snr_param}
\ee
We did not explore the parameter space too deeply, as the purpose of this work is to demonstrate that the correlations observed by W18 may be explained by hypothesis (i) or (ii) of the Introduction without resort to hypothesis (iii). There likely exists a set of parameters and a set of modifications to the assumed distributions $p(R_p|\a_\mu)$ above that do an even better job of reproducing the CKSM data. Despite this limited search, the parameters (\ref{eq:snr_param}) provide an excellent fit to the overall distribution of radii in the CKSM data. In particular, the mock planetary radius distributions for all the observed planets and for those with SNR$>20$ as well as the correlations between the radii of adjacent planets are nearly identical to the data. On average we find $64 \pm 2\%$ of planets with SNR$>20$ in agreement with $\sim 70\%$ in the CKSM data; the probability that the radius distributions for all observed planets and those with SNR$>20$ are identical to the respective CKSM distributions is $40 \pm 30 \%$; and the average Spearman correlation coefficient between the radii of adjacent planets $r_S = 0.57 \pm 0.04$ is also in complete agreement with the observed value of $0.56.$ In the four-component mock universe the correlation between the radii of the adjacent planets persists until $R_p \simeq 3.3 R_\earth$ (Figure \ref{fig:snr_spearmans}), which is also in agreement with the  CKSM data presented on Figure \ref{fig.pearsons}b. The average fraction of points on the $R_{p,j+1}$ vs. $R_{p,j}$ plot (with the W18 swapping criterion applied) falling into the box $(R_{p,j+1}, R_{p,j}) < 1 R_\earth$ is $0.02 \pm 0.01,$ the box  $1 R_\earth < (R_{p,j+1}, R_{p,j}) < 2 R_\earth$ is $0.26 \pm 0.2,$ the box $2 R_\earth < (R_{p,j+1}, R_{p,j}) < 4 R_\earth$ is $0.35 \pm 0.3,$ and the box $(R_{p,j+1}, R_{p,j}) > 4 R_\earth$ is $0.04\pm 0.1$ in good agreement with $0.04, \, 0.021, \, 0.29,$ and $0.04,$ respectively, for the CKSM data.

It is indeed easy to produce an ``observed mock" distribution of planetary radii that is similar to the one observed in CKSM (see Figure \ref{fig:snr.fit}ac) using the nine free parameters describing the systems $\a_a,$ $\a_b,$ $\a_c$ and $\a_d$. What is not easy or obvious is that we can reproduce all the following characteristics together -- the distribution of planetary radii, the distribution of planetary radii with SNR$> 20$, the correlation between the radii of adjacent planets, the visual appearance of such a distribution, and the Pearson and Spearman correlation coefficients as a function of the minimum radius $R_\mathrm{cut}$ of planets included in the correlation. The last of these characteristics is the hardest to reproduce. One may naively think that mere presence of the $\a_b$ and $\a_c$ systems, which predominantly have planets in the vicinity of $R_b$ and $R_c$, would give us a correlation between the radii of adjacent planets up to $R_\mathrm{cut} \simeq R_b.$ This however is not the case. Such a correlation extends only to $R_\mathrm{cut} \simeq R_a \simeq 1.4 R_\earth.$ The small-planet systems $\a_a$ have the intrinsic density $\sim R_p^{-3.2},$ which produces a considerably weaker correlation at smaller radii than the one observed in CKSM. The large-planet systems $\a_d$ have the intrinsic density $\sim R_p^{-1},$ which produces no significant correlation between the radii of adjacent plants at all.  It is nontrivial that the combination of $\a_a,$ $\a_b,$ $\a_c$ and $\a_d$ produces the entire suite of observations described at the start of this paragraph and shown in Figures \ref{fig:snr.fit} and \ref{fig:snr_spearmans}.

Note that sampling from the combined four-component distribution in the same way as we sampled in Section \ref{sec:corr1} and \ref{sec:boot} when considering hypothesis (i) would not produce as strong a correlation between the radii of adjacent planets as on Figure \ref{fig:snr.fit}bd. The results would look similar to Figures \ref{bootstrap}adcf.

In summary, we have demonstrated that a correlation between the radii of adjacent planets that is statistically indistinguishable from the correlation found by W18 can arise simply from assuming that planets are formed in systems of several distinct types.

%(3.2, 1.42, 0.12, 2.35, 0.23000000000000004, 3.5, 0.54, 0.3, 0.035, ' 0', 0.6357300884955752, 0.47932522123893806, 0.047002212389380536, 0.3959493376643575, 0.0, 0.0, 0.05673233654707669, 0.386566948493542, 0.0, 0.0, 0.573839483175307, 4.187619373362429e-41, 0.02403340014175302, 0.2642962321852357, 0.3483423900433371, 0.03988000962288673, 0.5735178774327605, 4.693289208184218e-41, 0.5282756858386014, 4.258488353068421e-32, 0.3284402843371684, 4.800933370584607e-09, 0.28129451171740777, 6.826463297735831e-05, 0.4984343834441776, 3.21455785248105e-07, 0.5421563147561352, 0.00034016539070072456, 0.19298829915570015, 0.33722243454416245, 0.07937295013141844, 0.4437732887511635, 0.06502529114072331, 0.3237841922922144, '0', 0.01859179765471604, 0.019891440657669335, 0.01686774800828868, 0.31022809302064147, 0.0, 0.0, 0.018911111152941976, 0.3241855306846594, 0.0, 0.0, 0.03834195435217268, 0.0, 0.008464040039612106, 0.024245172607768213, 0.027848344827785525, 0.01181861700116424, 0.03828304562990872, 0.0, 0.04281987849725813, 0.0, 0.06646029184729997, 0.0, 0.08526704192440379, 0.0, 0.09660864777200205, 0.0, 0.11733746057309029, 0.0, 0.20044759865862535, 0.0, 0.2650241222861332, 0.0, 0.2539915236182748, 0.0)

\bigskip

\section{Summary}\label{sec:sum}

We have shown that the apparent correlation between the radii of adjacent planets found by \cite{2018AJ....155...48W} can arise in two simple toy models, neither or which requires that planets ``know about" the properties of their neighbors. 

The first toy model correctly reproduces the correlation, although at the cost of an SNR distribution that is not consistent with the data and a radius distribution that is steeper than observed. Nevertheless, it shows that the correlation can arise in part from observational selection effects: in host systems that are difficult to observe---because of large stellar radii, photometric noise, or other system properties---all detected planets will have large radii, while in systems that are easy to observe most planets will have small radii close to the detection limit. This conclusion is similar to the argument of \cite{2019arXiv190702074Z}, although based on quite different methods. This hypothesis assumes that the planets ``don't know anything''; that is, that their probability of formation is guided by %requires 
a universal distribution function that depends only on the planetary radii. 

Our second toy model does not suffer from this shortcoming. This model is based on the hypothesis that planets ``know about the system they formed in'', that is, that the distribution of planetary radii may vary from system to system. With a simple version of this hypothesis requiring only four types of system we can reproduce the distribution of planetary radii observed in the CKSM catalog, the distribution of SNR, and even the correlation coefficients between the radii of adjacent planets.

We have also shown that the bootstrap simulations used by \cite{2018AJ....155...48W}, as well as an improved simulation that weights planets by their detectability, are statistically biased in that they seriously underestimate the occurrence rate of small planets, and therefore generate radius distributions that are less steep than the true distribution.  Therefore this method cannot be used either to derive the planetary radius distribution or to argue for the astrophysical nature of the correlation of the radii of adjacent planets.

W18 also found correlations in other properties of the planets in multi-planet systems, but we have not analyzed these findings. 

The solution to the bias problems that have plagued these analyses is to perform Bayesian analyses of the entire Kepler catalog, rather than just stars that have detected planets, and to include detailed models for the planet detection and vetting probability.

\cite{2017ApJ...849L..33M} have found similar correlations in the distribution of masses in a much smaller sample of {\it Kepler} multi-planet systems (37 systems and 89 planets). They concentrated on the correlation with respect to planetary masses, using masses derived from transit timing variations. In their tests they use a bootstrap algorithm that re-samples from the observed distribution, and thus is subject to the same criticisms as the bootstrap used in \cite{2018AJ....155...48W}; however, the main selection effects in their catalog arise from mutual inclinations and period ratios so the bias introduced by this method may be small.

\section*{Acknowledgements}
We are grateful to Kento Masuda, Erik Petigura, Lauren Weiss, Wei Zhu, and the anonymous referees for perceptive and useful comments. LM's support at the IAS is provided by the Friends of the Institute for Advanced Study.

\bibliography{PeasBibliography}{}
\bibliographystyle{aasjournal}

%\end{document}

\appendix
\section{Balanced bootstrap simulations in a catalog containing systems of different types}

% Let's assume we have $N$ systems. A fraction $k_\a$ of them belong to a type $\a$ and the rest $k_\b$ to a type $\b$ ($k_\a+k_\b=1$). There are two types of planets small $R_1$ and large $R_2.$ Systems $\a$ tend to have small $R_1$ planets with probability $p_1.$ Systems $\b$ tend to have large $R_2$ planets with probability $q_2.$ Large planets $R_2$ are always detectable. Small planets are detectable with probability $f_\a$ in the $\a$-systems and the probability $f_\b$ in the $\b$-systems.

% In the universe we have $N k_\a p_1$ small planets and $N k_\b q_2$ large planets. In the catalogue -- $N k_\a p_1 f_\a$ small planets and $N k_\b q_2$ large planets. Intrinsic fraction of small planets to large ones:
%\be 
%   r=\frac{N k_\a p_1}{N k_\b q_2}.
%\ee
% At the same time small planets would be detectable in $N k_\a f_\a + N k_\b f_\b$ of systems, if they could actually be there. Now let's try to reconstruct the intrinsic distribution using the accurate weighting:
%\be 
%    r_{est}=\frac{N k_\a p_1 f_\a}{N k_\b q_2} \frac{N}{N k_\a f_\a + N k_\b f_\b}= r \frac{f_\a}{k_\a f_\a + k_\b f_\b}=r \frac{1}{1-k_\b(1-f_\b/f_\a)}.
%\ee
% The estimated fraction coincides with the true one only if $f_\a=f_\b.$

% The above example is the simplest but not the easiest to take limits of. More complex version, but the one good for taking limits:

Let us assume we have a catalog containing $N$ stars that may or may not have detectable planetary systems. A fraction $k_\a$ of them belong to a type $\a$ and the rest $k_\b$ to a type $\b$ ($k_\a+k_\b=1$).
There are two types of planets, having radii $R_1$ and $R_2>R_1.$ Systems $\a$ have small $R_1$ planets with probability $p_1$ and large $R_2$ ones with probability $p_2.$ Systems $\b$ have large $R_2$ planets with probability $q_2$ and small ones with probability $q_1.$ Large planets $R_2$ are always detectable. Small planets are detectable with probability $f_\a$ in the $\a$-systems and probability $f_\b$ in the $\b$-systems.

In the universe we have $N k_\a p_1 + N k_\b q_1$ small planets and $N k_\b q_2+N k_\a p_2$ large planets. In the catalog we have $N k_\a p_1 f_\a +  N k_\b q_1 f_\b$ small planets and $N k_\a p_2+N k_\b q_2$ large planets. The intrinsic ratio of small planets to large ones is 
\be 
    r=\frac{k_\a p_1 + k_\b q_1}{k_\a p_2 + k_\b q_2}.
\ee
Small planets would be detectable in $N k_\a f_\a + N k_\b f_\b$ of systems. 
Now let us try to reconstruct the intrinsic ratio of small planets to large one using the weighting
\be 
    r_\mathrm{est}=\frac{\mbox{(number of small planets in the catalog)}}{\mbox{(number of large planets in the catalog)}}
    \times \frac{\mbox{(total number of systems observed)}}{\mbox{(number of systems in which small planets could be detected)}}.\nonumber
\ee
We have 
\be 
    r_{\mathrm{est}}=\frac{N k_\a p_1 f_\a + N k_\b q_1 f_\b}{N k_\b q_2 + N k_\a p_2} \frac{N}{N k_\a f_\a + N k_\b f_\b}
    = r \frac{k_\a p_1 f_\a + k_\b q_1 f_\b}{k_\a p_1 + k_\b q_1} \frac{1}{k_\a f_\a + k_\b f_\b}.
\ee    
The estimated ratio coincides with the true one only if $f_\a=f_\b$ or $p_1 = q_1.$
Otherwise there is no way to reconstruct the intrinsic occurrence unless we know which systems are type $\a$ and which are $\b.$ However it may not always be possible to distinguish systems $\a$ from systems $\b$ using observations of the current properties of the systems. There may be properties of the protoplanetary disk that were erased when the planets were formed and the star blew away the remainder of the disk.

\end{document}